%% file: main.tex
\newif\ifdouble
\newif\ifsingle
\newif\ifchange
\doubletrue

\documentclass[sigconf, usenames, dvipsnames]{acmart}

\usepackage{dblfloatfix} 
\usepackage{booktabs} 
\usepackage{arydshln}
\usepackage{subfig}
\usepackage{pdfpages}
\newcommand{\change}[1]{\textcolor{black}{#1}}

\AtBeginDocument{%
  \providecommand\BibTeX{{%
    \normalfont B\kern-0.5em{\scshape i\kern-0.25em b}\kern-0.8em\TeX}}}

\copyrightyear{2022} 
\acmYear{2022} 
\setcopyright{acmlicensed}\acmConference[UIST '22]{The 35th Annual ACM Symposium on User Interface Software and Technology}{October 29-November 2, 2022}{Bend, OR, USA}
\acmBooktitle{The 35th Annual ACM Symposium on User Interface Software and Technology (UIST '22), October 29-November 2, 2022, Bend, OR, USA}
\acmPrice{15.00}
\acmDOI{10.1145/3526113.3545702}
\acmISBN{978-1-4503-9320-1/22/10}

\newcommand{\system}{RealityTalk}

\begin{document}
\pagenumbering{arabic}
\pagestyle{plain}
\title{\system{}: Real-Time Speech-Driven Augmented Presentation for AR Live Storytelling}

\author{Jian Liao}
\affiliation{%
  \institution{University of Calgary}
  \city{Calgary}
  \country{Canada}}
\email{jian.liao1@ucalgary.ca}

\author{Adnan Karim}
\affiliation{%
  \institution{University of Calgary}
  \city{Calgary}
  \country{Canada}}
\email{adnan.karim@ucalgary.ca}

\author{Shivesh Jadon}
\affiliation{%
  \institution{University of Calgary}
  \city{Calgary}
  \country{Canada}}
\email{shivesh.jadon@ucalgary.ca}

\author{Rubaiat Habib Kazi}
\affiliation{%
  \institution{Adobe Research}
  \city{Seattle}
  \country{U.S.A.}}
\email{rhabib@adobe.com}

\author{Ryo Suzuki}
\affiliation{%
  \institution{University of Calgary}
  \city{Calgary}
  \country{Canada}}
\email{ryo.suzuki@ucalgary.ca}


\renewcommand{\shortauthors}{Liao, et al.}

\input{0-abstract}

\begin{teaserfigure}
\begin{center}
\includegraphics[width=1\textwidth]{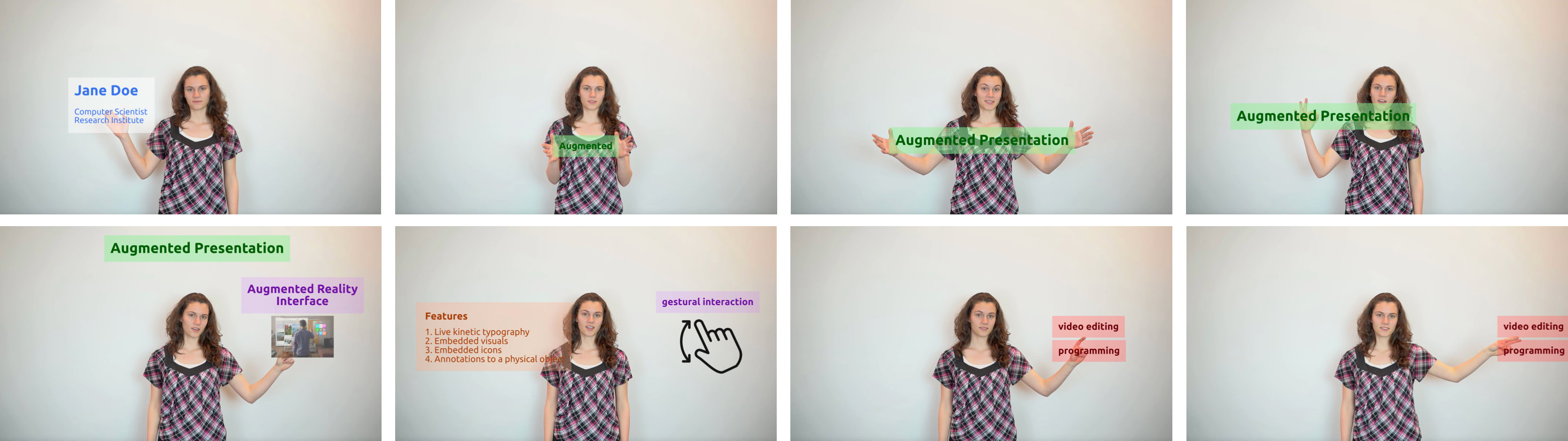}
\end{center}
\caption{\system{} augments live video presentations with \emph{speech-driven} interactions, where the live speech and supporting modalities (such as gestures) drive the triggering and manipulation of textual and graphical elements on the screen. 
}
\label{fig:teaser}
\end{teaserfigure}

\maketitle

\input{1-introduction}
\input{2-related-work}

\input{3-analysis}

\input{4-design-space}

\input{5-system-design}

\input{6-application}

\input{7-user-study}
\input{8-future-work}
\input{9-conclusion}

\begin{acks}
  This research was funded in part by the Natural Sciences and Engineering Research Council of Canada (NSERC), Adobe Gift Funding, and Snap Creative Challenge Research Award.
\end{acks}

\ifdouble
    \balance
\fi

\bibliographystyle{ACM-Reference-Format}
\bibliography{references}

\end{document}

%% file: 0-abstract.tex
\begin{abstract}
We present RealityTalk, a system that augments real-time live presentations with \textbf{\textit{speech-driven}} interactive virtual elements. Augmented presentations leverage embedded visuals and animation for engaging and expressive storytelling. However, existing tools for live presentations often lack interactivity and improvisation, while creating such effects in video editing tools require significant time and expertise. RealityTalk enables users to create live augmented presentations with real-time speech-driven interactions. The user can interactively prompt, move, and manipulate graphical elements through real-time speech and supporting modalities. Based on our analysis of 177 existing video-edited augmented presentations, we propose a novel set of interaction techniques and then incorporated them into RealityTalk. We evaluate our tool from a presenter's perspective to demonstrate the effectiveness of our system.
\end{abstract}

\ccsdesc[500]{Human-centered computing~Mixed / augmented reality}

\keywords{Augmented Reality; Mixed Reality; Augmented Presentation; Natural Language Processing; Gestural and Speech Input; Video}

%% file: 1-introduction.tex
\begin{figure*}[ht!]
\centering
\includegraphics[width=1\textwidth]{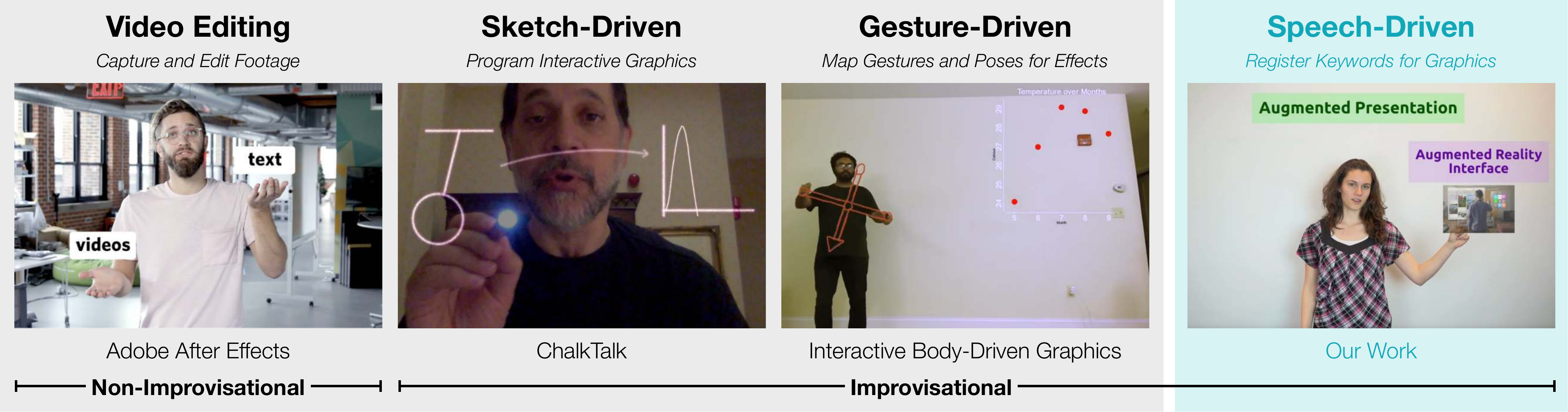}
\caption{Comparison from the video editing techniques~\cite{motiontracking} to improvisational augmented presentation, such as ``sketch-driven'' augmentation like ChalkTalk~\cite{perlin2018chalktalk} or ``gesture-driven'' augmentation like Interactive Body-Driven Graphics~\cite{saquib2019interactive}, our approach enables \textit{``speech-driven''} augmented presentation, which is a novel approach to improvisational and live augmented storytelling.
}
\label{fig:related-work}
\end{figure*}

\section{Introduction}
Augmented presentation --- a presentation practice that leverages embedded visuals and animations --- enables engaging, memorable, and expressive storytelling~\cite{saquib2019interactive, holoboard, perlin2018chalktalk}. 
For example, Hans Rosling's performance~\cite{joy-of-stats} uses interactive and embedded visualizations for engaging storytelling. Popular educational videos like BYJU's~\cite{BYJU} use video augmentations to seamlessly integrate graphics and visuals with the performance of the presenter. 
Due to their engagement and expressiveness, such augmentations are becoming more popular on the user-generated content platforms like YouTube or TikTok in the domain of education, advertisement, branding, and tutorials. 

Traditionally, such augmented presentations are mostly created with post-processing video-editing tools like Adobe After Effects, which requires significant time and expertise. Moreover, such post-production tools are not suitable for \textit{live} presentations and \textit{interactive} experiences. 
More recently, commercial presentation tools like Prezi Video~\cite{PreziVideo} and mmhmm.app~\cite{mmhmm} for live presentations facilitate seamless augmentation of live video feed with graphical elements. However, these tools enforce a \textit{scripted} and \textit{linear} presentation, thus lacking interactivity and improvisation which are necessary for semi-structured presentations.

The goal of this paper is to facilitate an interactive storytelling experience by enabling the presenter to prompt, embed, annotate, and manipulate textual and visual elements with \textbf{\textit{real-time speech-driven}} improvisational interactions (Figure~\ref{fig:teaser}). While prior works have explored live augmented presentations with gestures \cite{saquib2019interactive} and interactive sketching~\cite{perlin2018chalktalk, ryo2020reality}, \emph{speech-based} interaction techniques are relatively less explored in these settings (Figure~\ref{fig:related-work}).

To design our system, we have collected 177 videos from the existing video-edited augmented presentations, \change{which we have compiled as an \textbf{interactive gallery}}\footnote{\change{\url{https://ilab.ucalgary.ca/realitytalk}}}. 
Our analysis based on the collected videos suggests the importance of speech modality for a variety of interactions.
Based on the observed common augmentation techniques, we identify the design space for augmented presentations: 1) \textbf{textual} elements (e.g., title, keywords, and list), 2) \textbf{associated visual} elements (e.g., images, icons, and screens), 3) \textbf{location} of the elements (e.g., in front of the presenter, 3D surface, and physical object), 4) \textbf{interactions} with the elements (e.g., drag and drop, and scaling).

Based on the proposed interactions, we design and implement \system{}, a system that augments live presentations with speech-driven interactive virtual elements.
During the preparation/authoring phase, the presenter first registers the visual aids, such as images, icons, and videos by associating them with keywords through our authoring interface.
During presentation time, the user can interactively embed and manipulate these user-defined visual elements.

To achieve this, \system{} leverages the following components: WebSpeech API for speech recognition; Transformer-based natural language processing engine for real-time keyword extraction; MediaPipe for gesture recognition; React.js, Konva.js, and A-Frame for AR rendering of 2D/3D virtual elements that can be embedded in the real world.
We demonstrate our system for various presentation practices, such as business presentation, educational videos, and live e-commerce presentations.
We evaluate the usability of our system with 15 participants by asking them to author and perform their presentations.
Through the user study, we confirm that \system{} can benefit presenters to enhance the AR live storytelling performance through the real-time augmented presentation.

Finally, our contributions are
\begin{enumerate}
\item A novel approach for real-time augmented presentation based on \textit{speech-driven} interactive virtual elements.
\item Interaction techniques and design space identified through the analysis of the 177 videos from the existing video-edited augmented presentation.
\item A system implementation, application demonstration, and user evaluation of \system{}.
\end{enumerate}

%% file: 2-related-work.tex
\section{Related Work}

\subsection{Interactive Augmented Presentation Tools}
Traditionally, augmented presentation requires video editing and post-production, but recent works in HCI have started exploring \textit{interactive} augmented presentation in AR.
For example, HoloBoard~\cite{holoboard} presents a projection-based augmentation for educational scenarios.
ChalkTalk~\cite{perlin2018chalktalk} enables the creation of interactive visual elements with sketches in real-time.
In other works~\cite{Charade,novel-immersive-presentation,holostation,stein2012arcade,kim2018holobox}, these interactive augmented presentations are a powerful way of communicating and presenting ideas that can attract more audience's attention and engagement~\cite{saquib2019interactive, holoboard, perlin2018chalktalk}.
However, these existing tools still require a lot of preparation --- for example, HoloBoard requires the preparation of virtual objects and ChalkTalk only works with the prepared programmed elements.
To address this, the recent work in HCI has also explored an authoring tool that minimizes the user's preparation efforts.
For example, RealitySketch \cite{ryo2020reality} enables the creation of such an augmentation with simple sketches that can be used for augmented presentation in physics or mathematics classrooms. 
The most closely related to our work, interactive body-driven presentation~\cite{saquib2019interactive} allows the creation of such an augmented presentation with the user-defined mapping between the gestures and effects.
However, one of the limitations of their approach is \textit{``the mental overload of the presenter''}~\cite{saquib2019interactive} as the user needs to memorize what they should perform next.
In contrast, this paper contributes to a novel approach based on \textbf{\textit{speech-driven augmented presentation}} (Figure~\ref{fig:related-work}).
To the best of our knowledge, no existing works have explored this approach.

\subsection{Interactive Speech Interaction for AR}
Our system combines real-time speech recognition and natural language processing for augmented presentation. 
Such interactive speech interactions have been explored in AR interfaces. 
The use of speech-to-text in AR has been mostly used for accessibility applications~\cite{mosbah2006speech, peng2018speechbubbles, olwal2020wearable, tu2020conversational, miller2017use, schipper2017caption, findlater2019deaf, jain2018exploring, jain2015head, jain2018towards}.
For example, SpeechBubbles~\cite{peng2018speechbubbles} uses real-time speech transcription as textual aids for deaf and hard-of-hearing people. 
The speech-based keyword is often used to summon the command in AR. 
Hololens and other AR platforms often use speech-based keywords for a specific command and
CloudBits~\cite{muller2017cloudbits} explores the speech for a more natural and interactive interface for AR. 
In addition, StARe~\cite{rivu2020stare} has also explored the gaze and mid-air gestural interaction for such transcribed keywords in real-time for AR interactions.
Combined gesture and speech have been also explored in~\cite{piumsomboon2014grasp} and~\cite{irawati2006evaluation}
However, these prior works mostly focus on \textit{conversation} and have not explored these speech-based augmentations for \textit{presentation} purpose.
Taking inspiration from these works, we explore how we can combine speech and gestural interaction for real-time augmented presentation.

\subsection{Body-Driven Embodied Interaction}
\change{In addition to live speech, our system also incorporates gesture as the supporting modality to create engaging augmented presentations.} 
Recent developments in computer vision and depth camera (e.g., OpenPose~\cite{cao2017realtime} and Microsoft Kinect) have drawn increased attention to how embodied interaction~\cite{dourish2004action} promotes increased engagement, social interactions, and playful experiences~\cite{kang2016sharedphys}. 
Within body-driven embodied interaction, Saquib et al.~\cite{saquib2019interactive} presented their work using a Kinect-based interface for mapping a variety of body movements to a wide range of graphical manipulations.
Researchers in HCI have explored a variety of novel approaches for human gestures~\cite{krueger1985videoplace,steins2013imaginary,yan2018virtualgrasp}, which are also compiled as gesture taxonomies in HCI~\cite{aigner2012understanding,steins2013imaginary,baudel1993charade}. In this paper, we also incorporate gesture interactions for augmented presentation by leveraging a webcam and computer vision (e.g., MediaPipe~\cite{MediaPipe}).

\subsection{Authoring Tools for AR}
Real-time authoring allows the user to create expressive virtual elements that respond in real-time, as opposed to creating and crafting graphical effects in prototyping tools~\cite{pronto}). RealitySketch~\cite{ryo2020reality} presented an augmented reality interface for sketching interactive graphics and visualizations, which embeds responsive graphics in AR through real-time authoring with dynamic sketching. In this paper, we consider how to leverage real-time authoring to enhance real-time video presentation through holistic embodied interaction in performance time. In this vein, prior systems in HCI, such as Rapido~\cite{rapido} and ChalkTalk~\cite{perlin2018chalktalk}, allow the creation of interactive AR prototypes using demonstrated examples and interactive mapping of graphical effects during performance time. 
This paper explores a natural and organic real-time authoring approach for augmented presentation in performance time, similar to ChalkTalk~\cite{perlin2018chalktalk}.

%% file: 3-analysis.tex
\section{Analysis of Existing Augmented Presentations}

To better understand the common interaction techniques used in augmented presentation, we analyzed a set of 177 existing augmented presentation videos. 

\begin{figure*}[ht!]
\includegraphics[width=\textwidth]{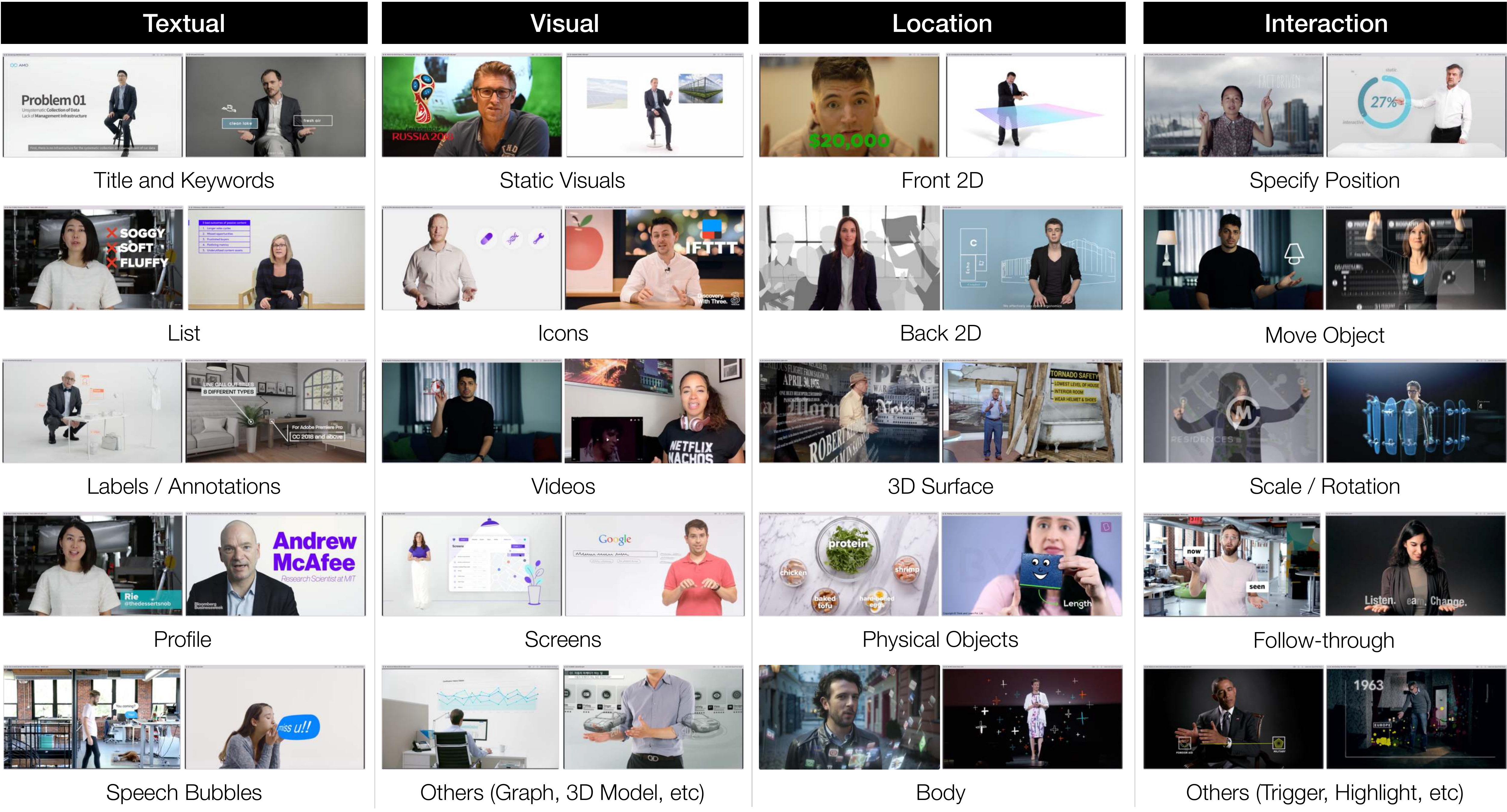}
\caption{Design Space Analysis -- We collected a set of 177 existing augmented presentation videos and observed the interaction design patterns in four dimensions. Each dimension has five categories, which are the commonly used techniques within the augmented presentation. \change{All materials and an interactive gallery are available at \url{https://ilab.ucalgary.ca/realitytalk}}}
\label{fig:design-space}
\end{figure*}

\subsection{Corpus, Methodology, and Initial Findings}
\subsubsection{Dataset}
To collect the videos, the authors manually searched examples from popular video platforms (e.g., YouTube and Vimeo).
We first tried to collect with a systematic search result based on a specific keyword like \textit{``augmented presentation''}, but we did not obtain meaningful results mostly because 1) there is no common keyword associated with these augmented presentation to date, 2) example videos or augmentations are often used for different context or purposes.
Therefore, we collected these videos mostly based on manual visual search using visual search platforms like Pinterest or Google Image Search. 
Because of this, we do not argue that our corpus is the systematically collected examples nor the representative examples for the augmented presentation. 
Our goal of this analysis is rather to identify the common techniques for augmented presentation and share the initial insights to the HCI community.

\subsubsection{Methodology}
We first collected 538 video augmentation videos available on the Internet, then filter out these videos by focusing only on the presentation or storytelling purposes.
After the filtering process, we have identified 177 videos for our analysis.
Given the 177 collected videos, two of the authors conducted open-coding to identify a first approximation of the dimensions and categories.
Next, all authors reflected upon the initial design space to discuss the consistency and comprehensiveness of the categorization.
After this process, two authors performed systematic coding with individual tagging for the categorization of the complete dataset.
Finally, we reflected upon the individual tagging to resolve the discrepancies to obtain the final coding results.

\subsubsection{Initial Findings and Design Rationale}
Our initial findings with the collected augmented presentations suggest the presenter's \textit{live narration} plays a critical role in triggering a wide range of visual and textual elements as aids. 
Therefore, in our system, we decided to explore this \textbf{\textit{speech-based augmentation}} as the primary modality for presenters to augment their presentations, and create our authoring tool and interaction techniques centered around the spoken keywords during the presentation time.

%% file: 4-design-space.tex
\section{Design Space of Speech-based Augmentation}
We have identified the design space as the following four dimensions: 1) textual elements, 2) visual elements, 3) location of the elements, and 4) interaction with the elements (Figure~\ref{fig:design-space}). 

\newcommand{\subsub}[1]{\noindent\textbf{\textit{- #1:}}}

\subsection{Textual Elements}
\subsub{Title and Keywords}
Title and keywords are simple text transcriptions for prompting the important information.
In most of the videos, the simple text transcription helps the audience see and memorize what the presenter says.

\subsub{List}
Many augmented presentations also use the list structure to show keywords. 
These lists are often associated with a specific numbering spoken term, such as ``First'', ``Second'', and ``Third''. 

\subsub{Label and Annotation}
Label and annotation are the textual element associated to a specific location or physical object, which is often connected with a line or an arrow to indicate the association.

\subsub{Profile}
Another commonly used text caption is displaying the profile description. 
When the presenters introduce themselves, the profile caption shows up with their names and affiliations. 

\subsub{Speech Bubbles}
Speech bubbles are used for conversational textual elements. For example, speech bubbles are commonly seen in displaying conversations (e.g., text messages) on the phone.

\subsection{Visual Elements}

\subsub{Static Visuals}
The augmented presentation videos often leverage associated visuals
The most frequently used one is static visuals like images and pictures, which show up along with keywords or replace the keywords with visual information.
Such visual aids help the audience understand and memorize the information.

\subsub{Icons}
Similar to pictures and images, icons are also commonly used for associated visuals. In contrast to the static visuals, icons are used for a simple representation of the associated keyword.
Such icons are used when describing some abstract or conceptual keywords.

\subsub{Videos}
Some augmented presentations use videos as a visual aid. For example, reaction video, as one of the popular genres of online videos, often uses embedded video while presenting their reactions.

\subsub{Screens}
The presenter also embeds the interactive screen like website to display information.
These embedded screens would be an important approach to show the synchronous user interactions with existing user interfaces. 

\subsub{Other Visuals}
Other visuals such as graphs, drawings, and 3D models are also used in some augmented presentations. For example, Hans Rosling's the joy of stats~\cite{joy-of-stats} presentation video uses various embedded graphs to display the statistical information.

\begin{figure*}[h!]
\centering
\includegraphics[width=0.9\linewidth]{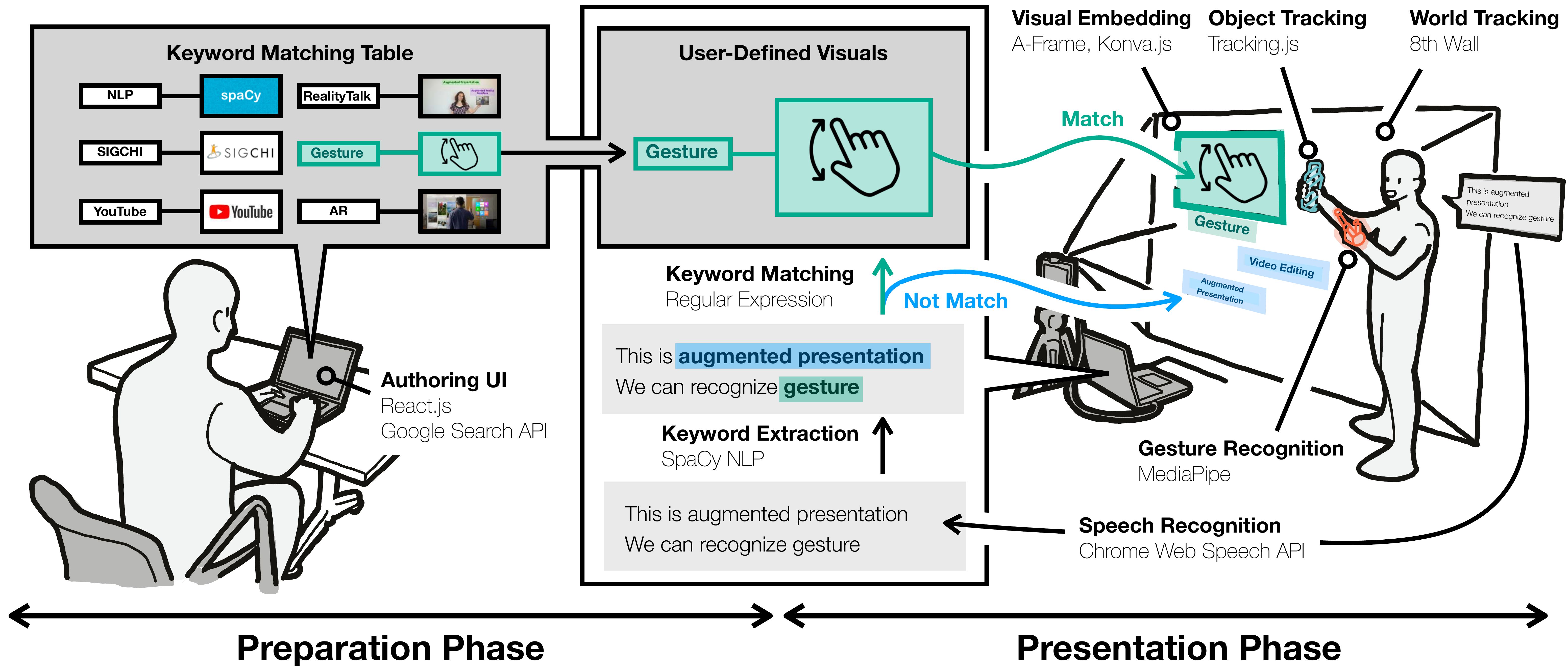}
\caption{System Overview -- \system{} offers both preparation and presentation features. The system uses a transformer-based NLP engine for keyword extraction and a simple keyword matching for associated visual elements.}
\label{fig:system-overview}
\end{figure*}

\subsection{Location of the Elements}

\subsub{Front 2D Embed}
Presenters often specify the location where these textual and visual elements show up during the presentation.
For example, front 2D elements are simply embedded through frontal immersion, which are spatially in front of the presenter.

\subsub{Back 2D Embed}
Back 2D elements are embedded behind the presenter, which normally utilizes body segmentation and masking to partially hide the elements that are overlapped with the presenter.

\subsub{3D Surface}
Alternatively, the text is also rendered in 3D space, such as vertical or horizontal surface. 
This gives a more immersive effect to the presentation by adding a 3D and depth effect.

\subsub{Physical Objects}
In some use scenarios, we have also identified texts and labels that are attached to physical objects. 
This is useful when the presenter explains something based on the physical object, such as explaining the ingredients and recipe for a cooking video.

\subsub{Body}
Finally, the text information is attached based on the body parts. 
Most commonly observed examples are the text is shown up on the hand or finger. 
We also observed that these virtual elements can be shown up around the face or body.

\subsection{Interaction with the Elements}

\subsub{Specify the Appearance Position}
Gestural interactions~\cite{aigner2012understanding,wobbrock2009user-defined,gesturesinar} are also common in augmented presentations.
For example, the presenter points out the position with her index finger to show up a title or keyword. 
This also includes pointing out the physical object to bind the label. 

\subsub{Moving Object}
Alternatively, the gestural interaction is also used to directly manipulate the object, such as dragging and dropping the textual or other visual elements.
This also includes removing virtual elements outside the screen through a throwing away gesture. 

\subsub{Scale and Rotation}
We also observed that the presenter often uses gestures for scaling the text.
This can be used to create an interactive animation with the text. 
We rarely observe the rotating gesture in the presentation, but this is also another interaction. 

\subsub{Follow-through}
Follow-through is commonly used within augmented presentation, which is essentially binding the textual and visual elements with hand motions.

\subsub{Other Interactions}
The other interactions can include animation triggers, highlighting the visual elements, interacting with embedded screens.

%% file: 5-system-design.tex
\section{\system{}: System Design}

\system{} is a system designed for speech-drive real-time augmented presentation.
Given the above design space exploration, \system{} will embed the interactive textual and visual elements based on the speech. 
This facilitates the improvisation and flexibility often required for semi-structured presentations and performances.

\subsection{Overview and Basic Setup}

\subsubsection{Overview}
\system{} incorporates the following workflow:

\subsub{Preparation Phase} the presenter associates the prepared visuals such as images, icons, and graphs with the keyword through our simple authoring interface.

\subsub{Presentation Phase} the presenter can show the interactive textual and visual elements based on the real-time speech in AR.

\ \\ \noindent
To achieve this, \system{} leverages the following components:

\subsub{Speech Recognition} Recognize and transcribe the speech into text in real-time with WebSpeech API. 

\subsub{Keyword Extraction} Extract, filter, and detect keywords based on a Transformer-based Natural Language Processing Engine. 

\subsub{Keyword Matching} Based on the user-defined keywords during the preparation phase, the system matches and associates the keyword with a visual.

\subsub{Visual Embedding} Embed interactive visual and textual elements based on WebXR.

\subsub{Gesture Recognition} Recognize the hand and body pose based on MediaPipe for real-time gestural interaction.

\subsub{World and Object Tracking} Detect and track 3D environment with 8th Wall and physical object by leveraging a simple computer vision based color tracking.

\subsubsection{Basic Setup}
The goal of our system is to democratize the augmented presentation for non-expert users for many situations with minimal effort (e.g., school teachers, business persons, video creators, children, etc.).
Thus, we intentionally design \system{} to be a system that can run on any commodity machines that have a webcam, running on a web browser. 
The system will show real-time augmentation effects while the presenter is giving a presentation.


\subsection{Simple Authoring User Interface}

By default, the system shows animated textual elements without any user specification.
For example, in Figure~\ref{fig:teaser}, the user can show the augmented textual elements such as \textit{``augmented presentation''}, \textit{``video editing''}, and \textit{``programming''}, without any preparation or authoring.
However, when the user wants to add more visual aids associated with the keyword, the user then needs to specify keywords and associated contents.
For example, in Figure~\ref{fig:teaser}, the user can embed images for \textit{``augmented reality interfaces''} or icons for \textit{``gestural interaction''}, the user can further enrich the presentation through these associated visual aids.

\subsubsection{Supported Embedded Visual Elements}
During the preparation phase, the presenter can create a list of keywords and associated images or visuals that the presenter want to embed.
There are a couple of visual elements that the presenter can add, which include: 1) images, 2) icons, 3) videos, and 4) embedded screens (website).
Through a simple preparation phase (described below), the user can add this associated information, which can be automatically triggered when the user speaks.

\subsubsection{A Simple Keyword Matching Table Approach}
\system{} provides a simple authoring user interface based on the keyword matching table. 
The authoring interface has two columns similar to an Excel spreadsheet: one is the user-specified keyword, and another one is the associated visual elements. 
When the user adds a keyword, then the presenter can add any visuals, such as images, icons, and videos in the right column, based on the URL. 
To help the user easily find the associated images, the system also automatically shows possible images and videos extracted from Google Search API, so that the user can quickly add these images and videos to reduce the preparation time. 
When the presenter specifies an embed URL, then the system also embeds the interactive web browser based on iframe during the presentation. 
Other embed information such as tweets, maps, or interactive HTML can also be embedded in the same way. 

\subsubsection{Design Decision and Rationale}
To embed visuals associated with keywords, we originally tried an automated approach based on the heuristic image search, such as showing the top image obtained from the Google Image search. 
However, we soon realized that this approach does not perform well because of the randomness of the associated images. 
These automated visual aids are rather distracting to the presentation, increasing the noise of the presentation and sometimes making the presenter annoyed due to the lack of control.
On the other hand, we have also learned that the sophisticated and highly customizable authoring interface can easily overwhelm the user, enforcing the user to focus on the preparation rather than improvisational storytelling. 
Through the pilot user testing, we have concluded this simple authoring interface can minimize the presenter's efforts, while it can effectively augment and improve the presentation.

\subsection{Real-Time Augmented Presentation}
After the preparation phase, the user can now perform a live presentation using \system{}.

\subsubsection{Textual Elements}
\begin{figure}[h!]
\centering
\includegraphics[width=0.49\linewidth]{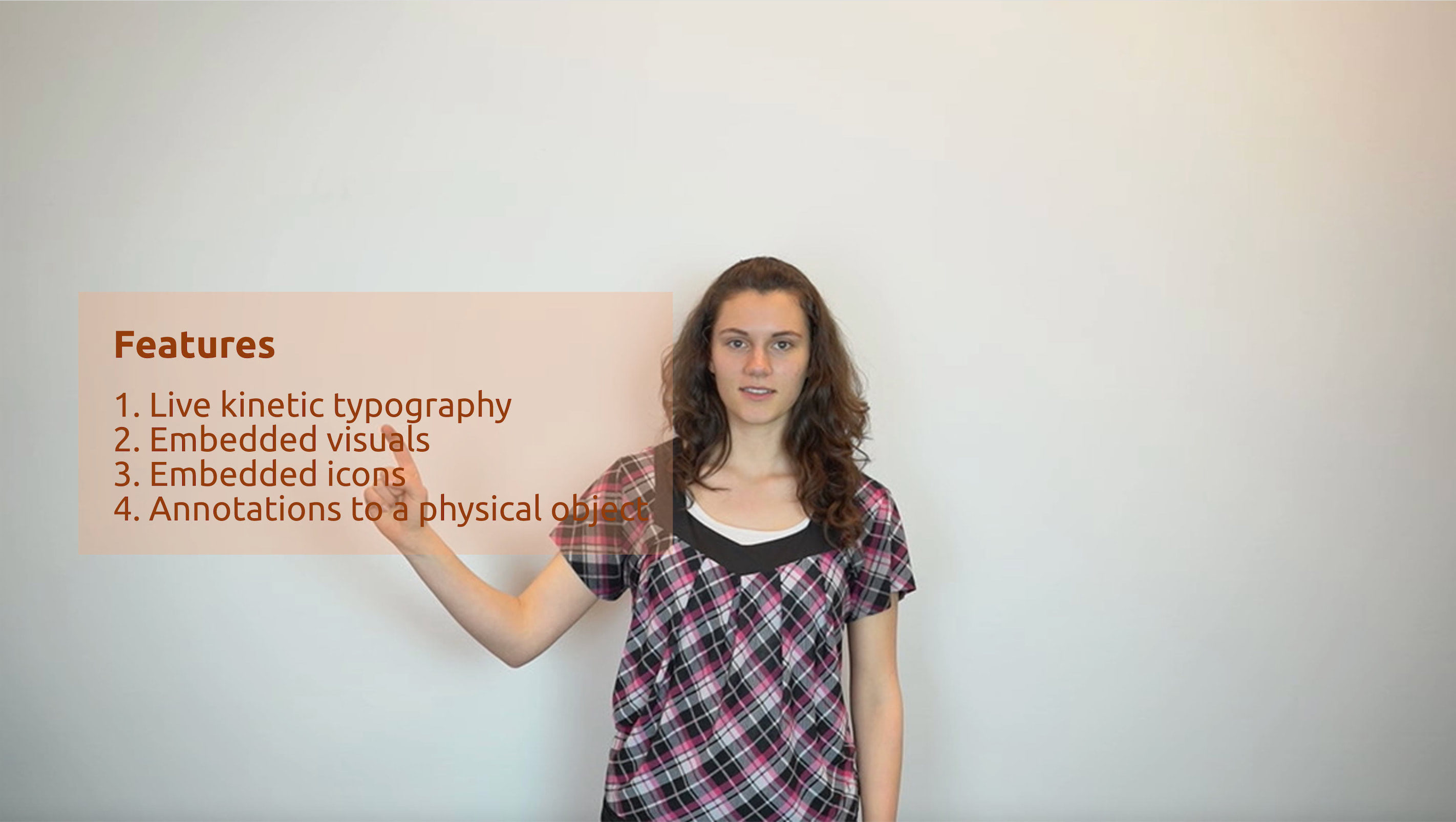}
\includegraphics[width=0.49\linewidth]{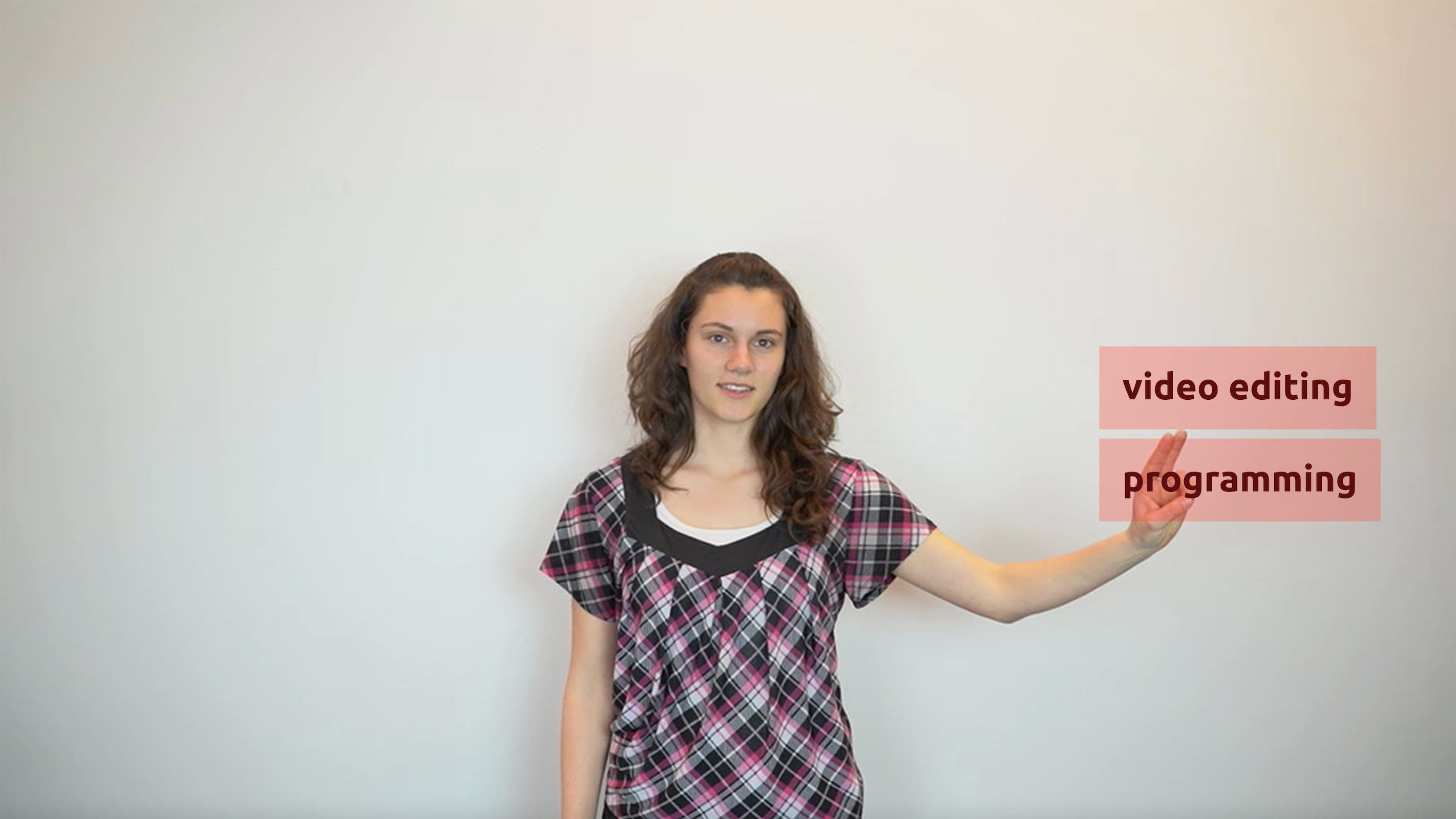}
\caption{Textual Elements -- A front 2D embedded list with the location specified by the pointing gesture and keywords that are moved away by the presenter.}
\label{fig:textual}
\end{figure}

First, the system can embed transcribed textual elements.
The system can show the animated speech caption based on real-time speech recognition and keyword extraction. 
As Figure~\ref{fig:textual} shows, when the presenter speaks, the extracted keywords are automatically shown up on the screen, similar to live transcripts.
In contrast to showing all the transcribed text in the presentation, the system only shows the important keywords for the live transcript.
We use NLP engine for the keyword extraction, which can also detect compound words like ``Human Computer Interaction'', as opposed to ``Human'', ``Computer'', and ``Interaction''. 

\change{Informed by the design space analysis, the system also supports a simple template behavior specifically for \textit{list} and \textit{profile}.}  
For example, when the presenter explains something with a bullet list in the presentation, the presenter usually says ``First'', ``Second'', ``Third''. With the default animated textual elements, the textual elements would get unorganized and couldn't deliver an enumerated behavior. Thus, the system supports the ``list mode'', when these pre-defined keywords are shown as a list (Figure~\ref{fig:textual} Left). 
Alternatively, when the presenter introduces themselves by saying ``my name is John'', the system also supports the ``profile mode'' to create a template profile showing information about the presenter.

\begin{figure}[h!]
\centering
\includegraphics[width=0.49\linewidth]{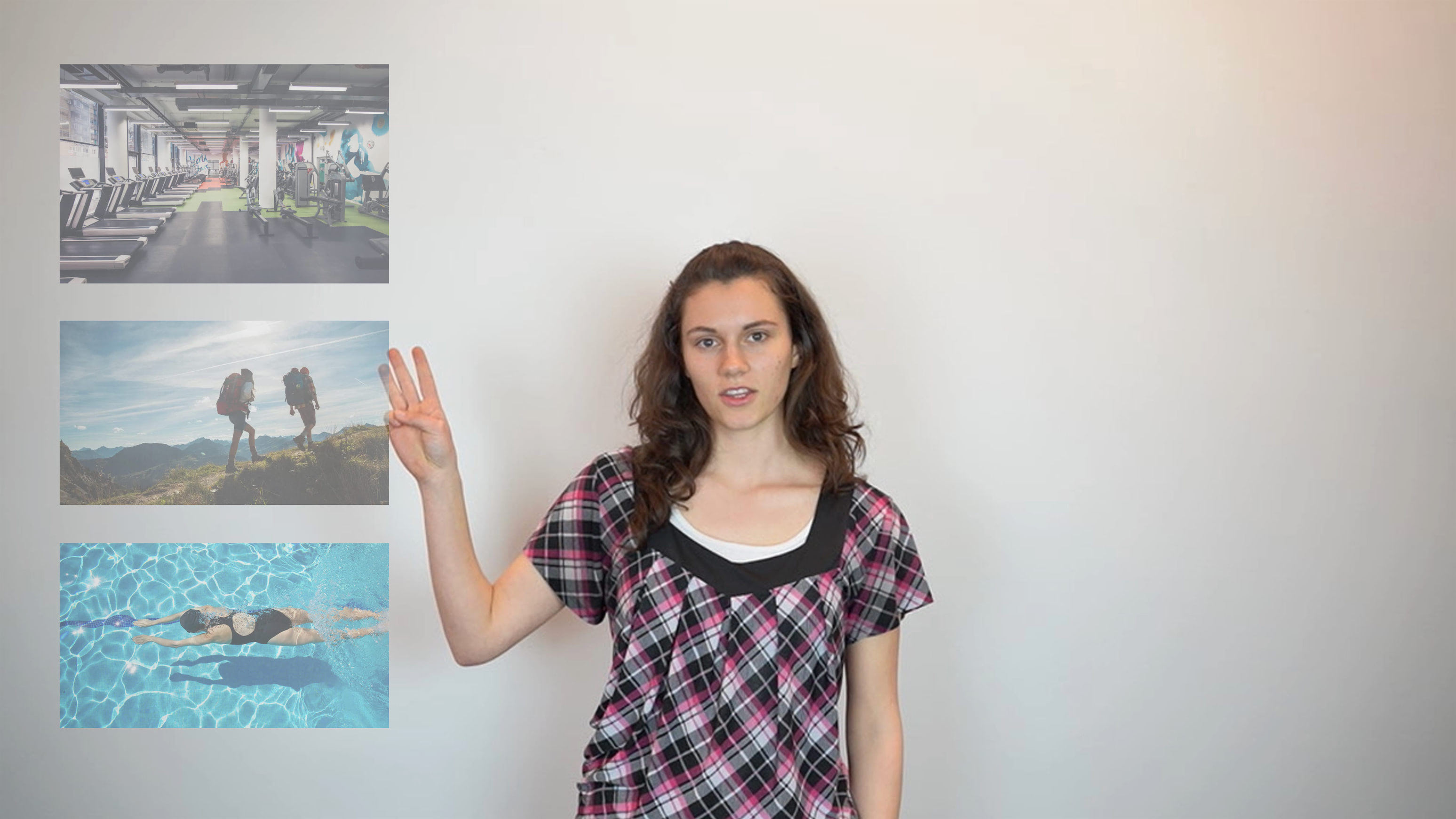}
\includegraphics[width=0.49\linewidth]{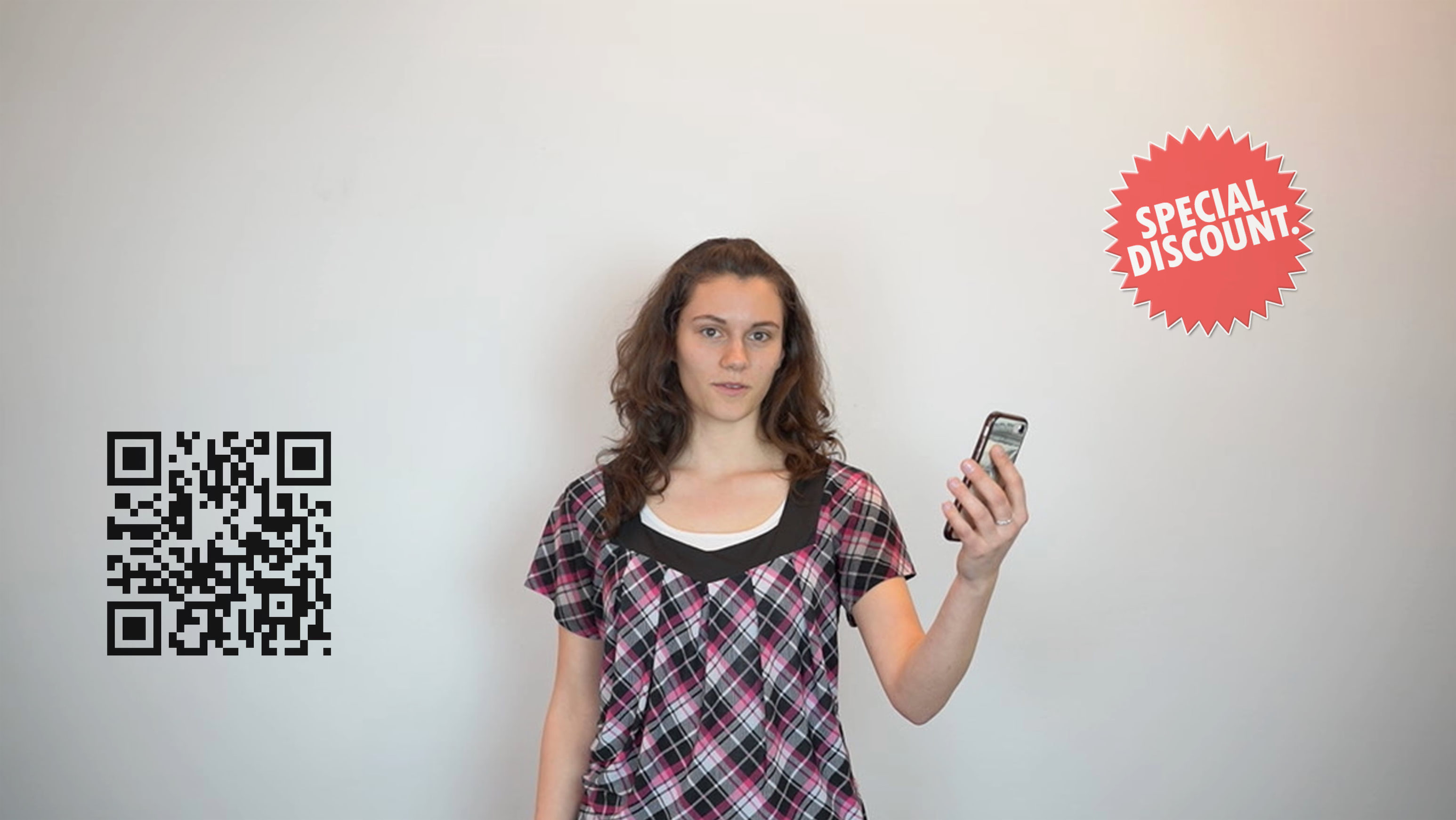}
\caption{Embedded Visuals -- Embedded static visuals that are triggered by speech input through keyword matching, demonstrating improvisational use cases in an e-commerce live-streaming setting.}
\label{fig:visual}
\end{figure}

By default, these textual elements will disappear with a certain duration (4 seconds), but the user-defined keywords can stay longer (10 seconds). The user can also manually specify the duration in the authoring user interface.
The textual elements have a transparent background, and the color of the text and background is changed randomly by default. 
To avoid unnecessary overlap between multiple embedded texts or between the text and the user's face, these textual elements are shown on the side of the center of the screen with a certain offset from the previous text, while the user can also specify the location of the text through the gestural interaction (described in the Supporting Modalities section).

\subsubsection{Embedded Visual Elements}
Based on the user-defined keywords, the presenter can further augment the presentation based on the prepared or improvisational associated visual information.
As we discussed in the authoring tool section, the system supports images, icons, videos, and embedded screens as the associated visual aids. 
For example, Figure~\ref{fig:visual} illustrates the presenter can show three photos based on the keywords or show a QR code based on a user-defined trigger word.
These visual elements are also interactively manipulated with gestural interaction (described below) so that the user can make it more interactive. 

\begin{figure}[h!]
\centering
\includegraphics[width=0.49\linewidth]{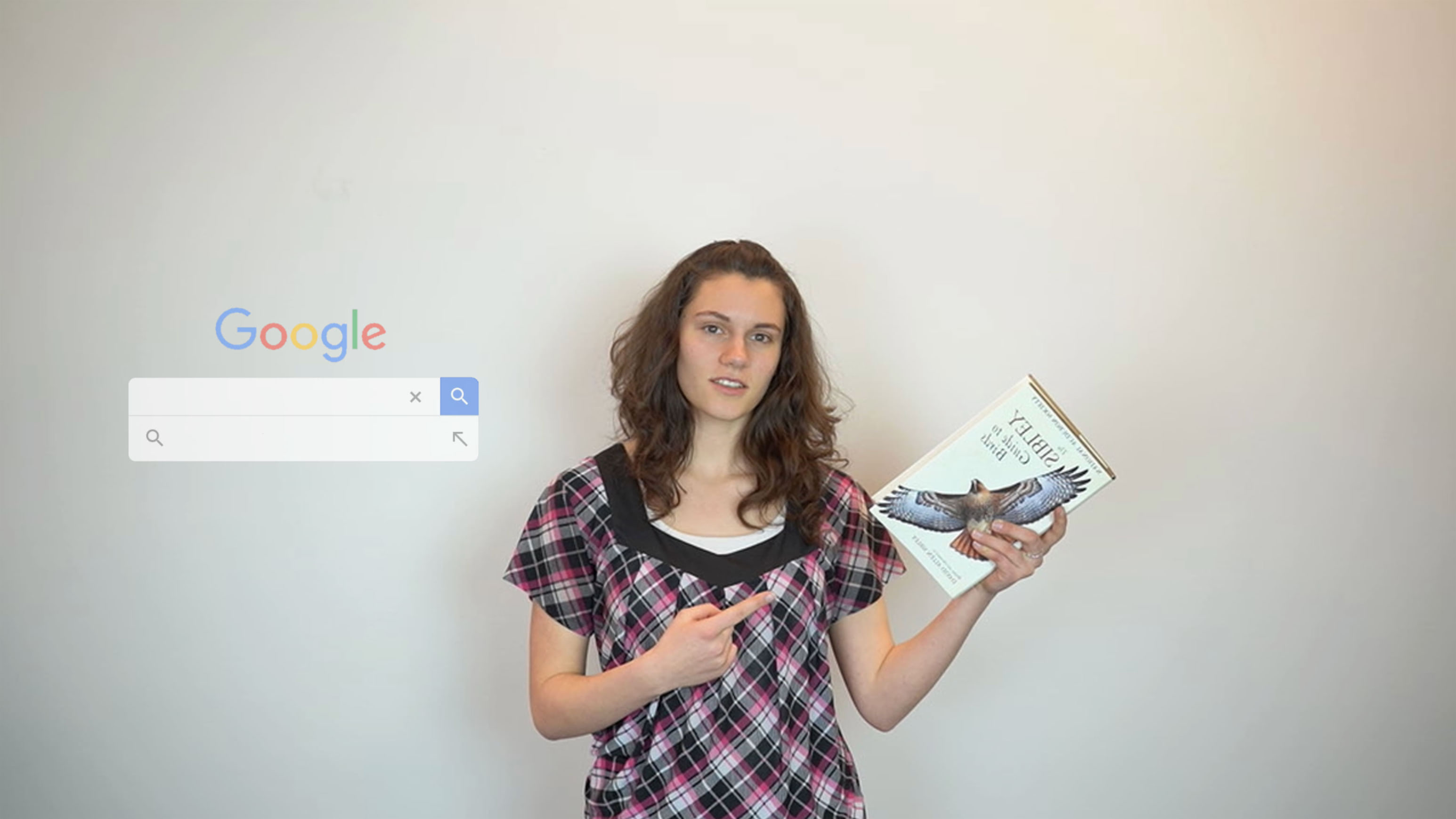}
\includegraphics[width=0.49\linewidth]{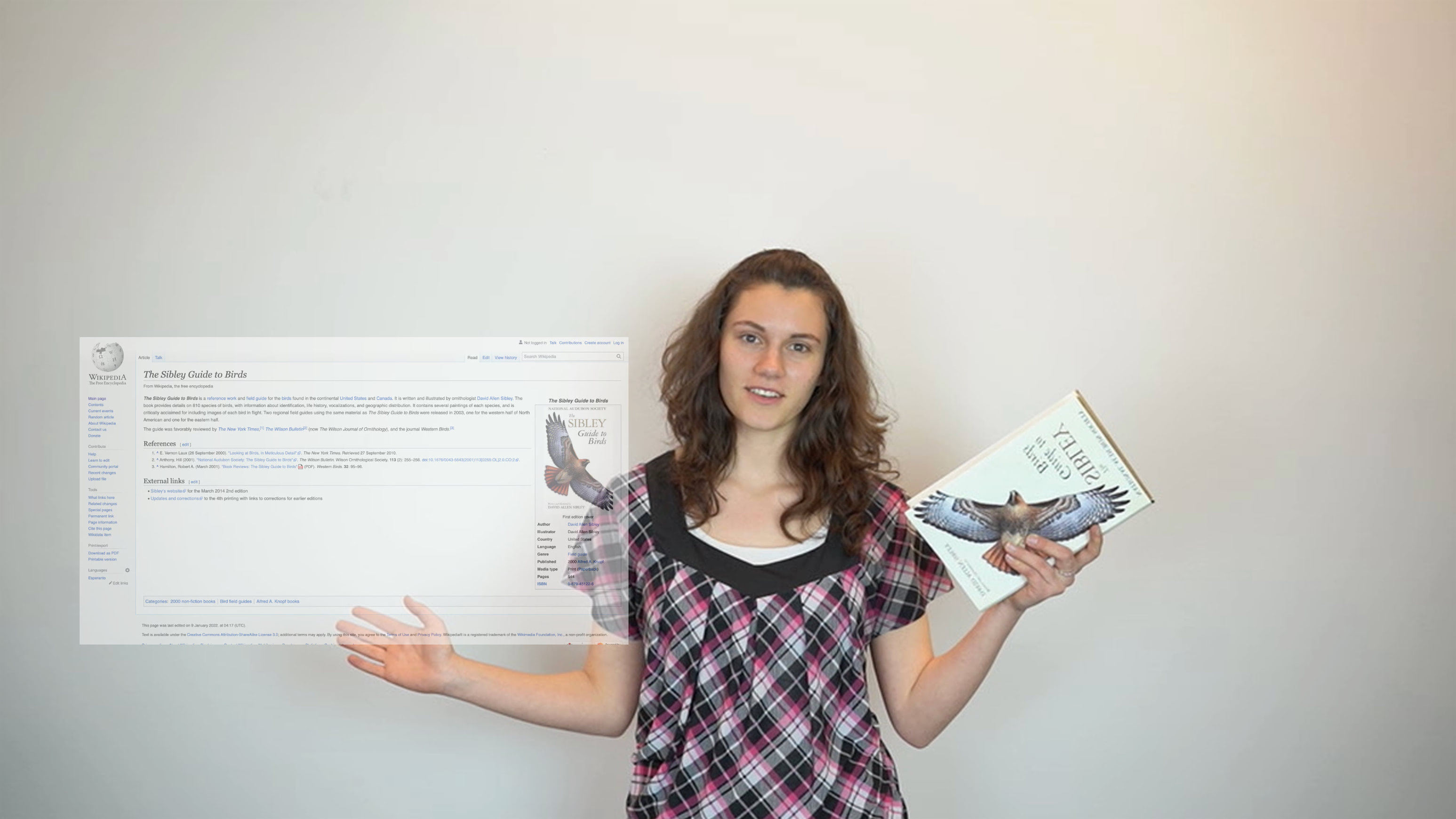}
\caption{Embedded Screens -- Embed an existing user interface \change{with iframe and html2canvas} into a presentation, which could allow the presenter to perform a Google search and show the associated Wikipedia page during the presentation.}
\label{fig:embedded-screen}
\end{figure}

Figure~\ref{fig:embedded-screen} also illustrates how the user can also add the screens based on the specified URL. For example, the user can show a Google or Wikipedia screen as the embedded iframe element. 
All of these visual elements are by default semi-transparent so that they create a more immersive effect. 
The video can be embedded as an iframe with the embed URL from YouTube or Vimeo, in which the video automatically plays without sound when shown in the presentation. 
Since the system cannot change the opacity of iframe, only the embedded screen is not transparent for the visual elements. 
When showing the visual, the user can also specify whether the keyword should be also shown or not, so that the user can only show the visuals without associated transcribed keywords. 
These visual stays on the screens for 10 seconds by default, but the user can also specify the duration of the appearance time of each visual.

\subsubsection{Supporting Modalities: Hand Tracking and Virtual Object Manipulation}
The interactivity with the embedded elements can greatly enhance the expressiveness and engagement of the augmented presentation. 
The system allows the user to interact with both textual and visual elements interactively through gestures. 

\begin{figure}[h!]
\centering
\includegraphics[width=0.49\linewidth]{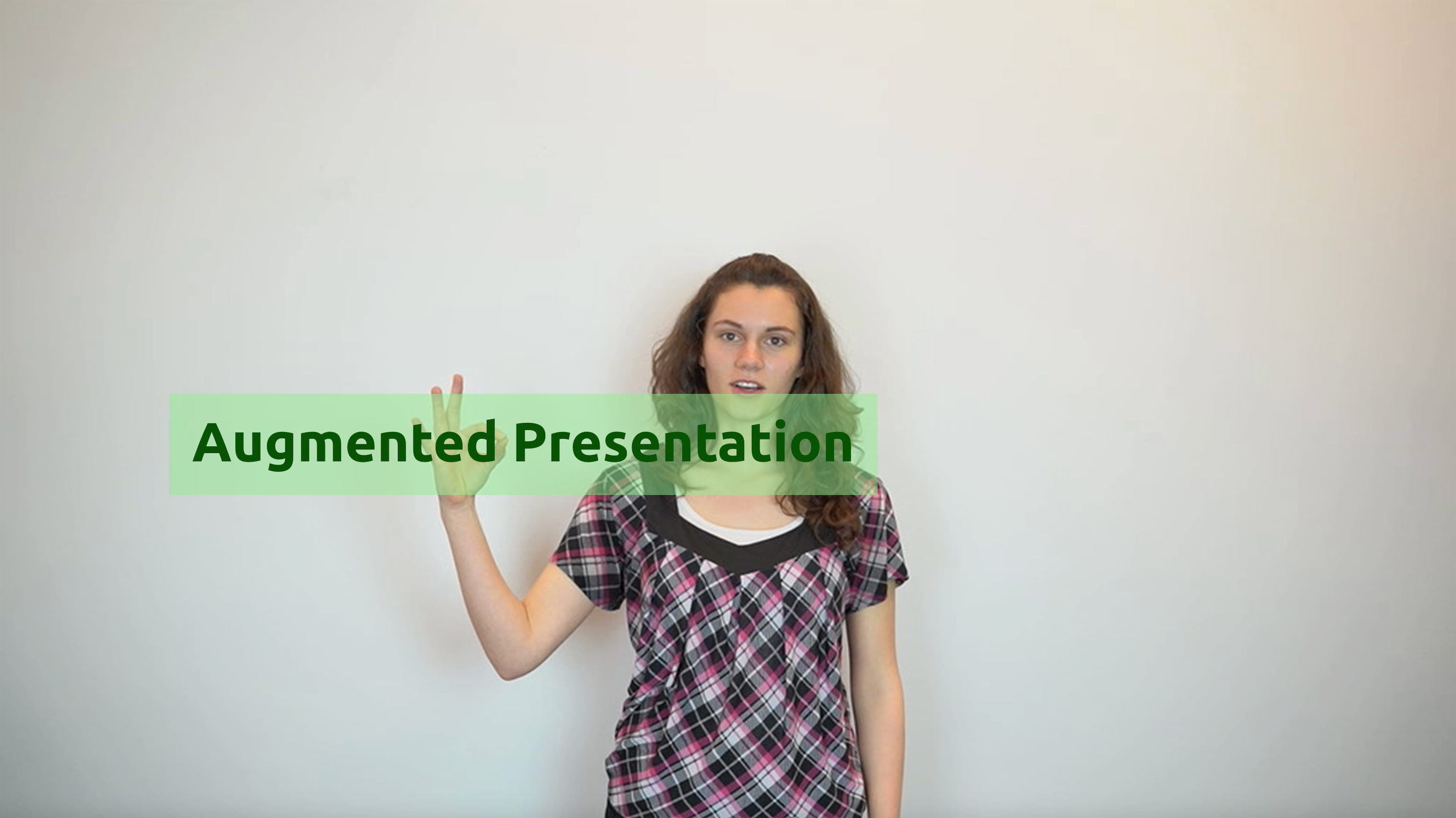}
\includegraphics[width=0.49\linewidth]{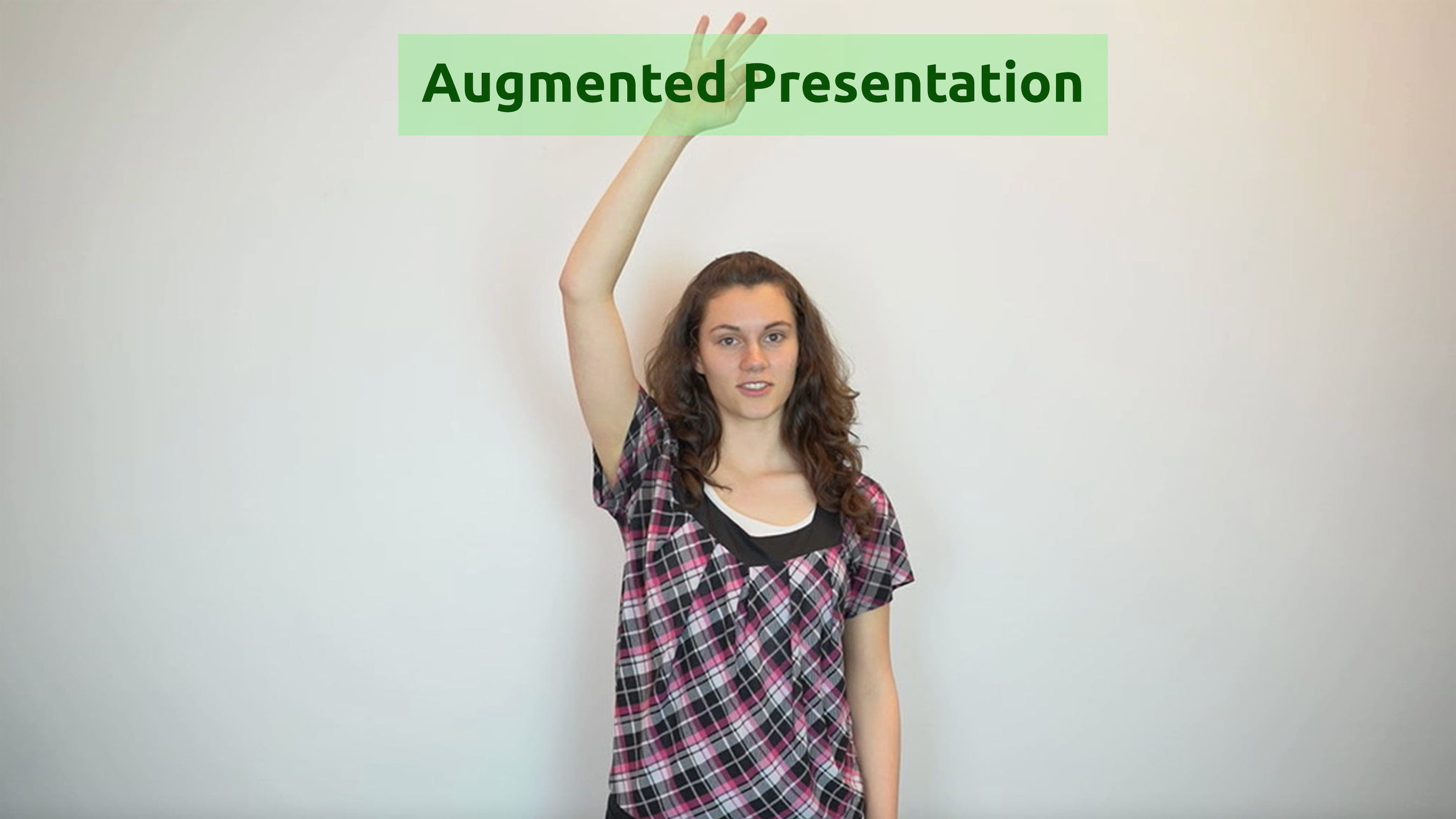}
\caption{Gestural interaction for moving an object through finger pinching gesture.}
\label{fig:gesture-moving}
\end{figure}

Informed by the design space analysis of augmented presentation videos, the system provides the following set of gesture manipulation: 
1) \textbf{pointing} the embedded position for the next element,
2) \textbf{dragging and dropping} the existing elements,
3) \textbf{scaling and rotating} the existing elements,
4) \textbf{removing} the existing elements.
For each of these gestures, the user can perform the following gestures: 1) pointing out with the index finger, 2) pinching gesture with the index finger and thumb of one hand, 3) pinching gesture with two hands, 4) swiping gestures. 

\begin{figure}[h!]
\centering
\includegraphics[width=0.49\linewidth]{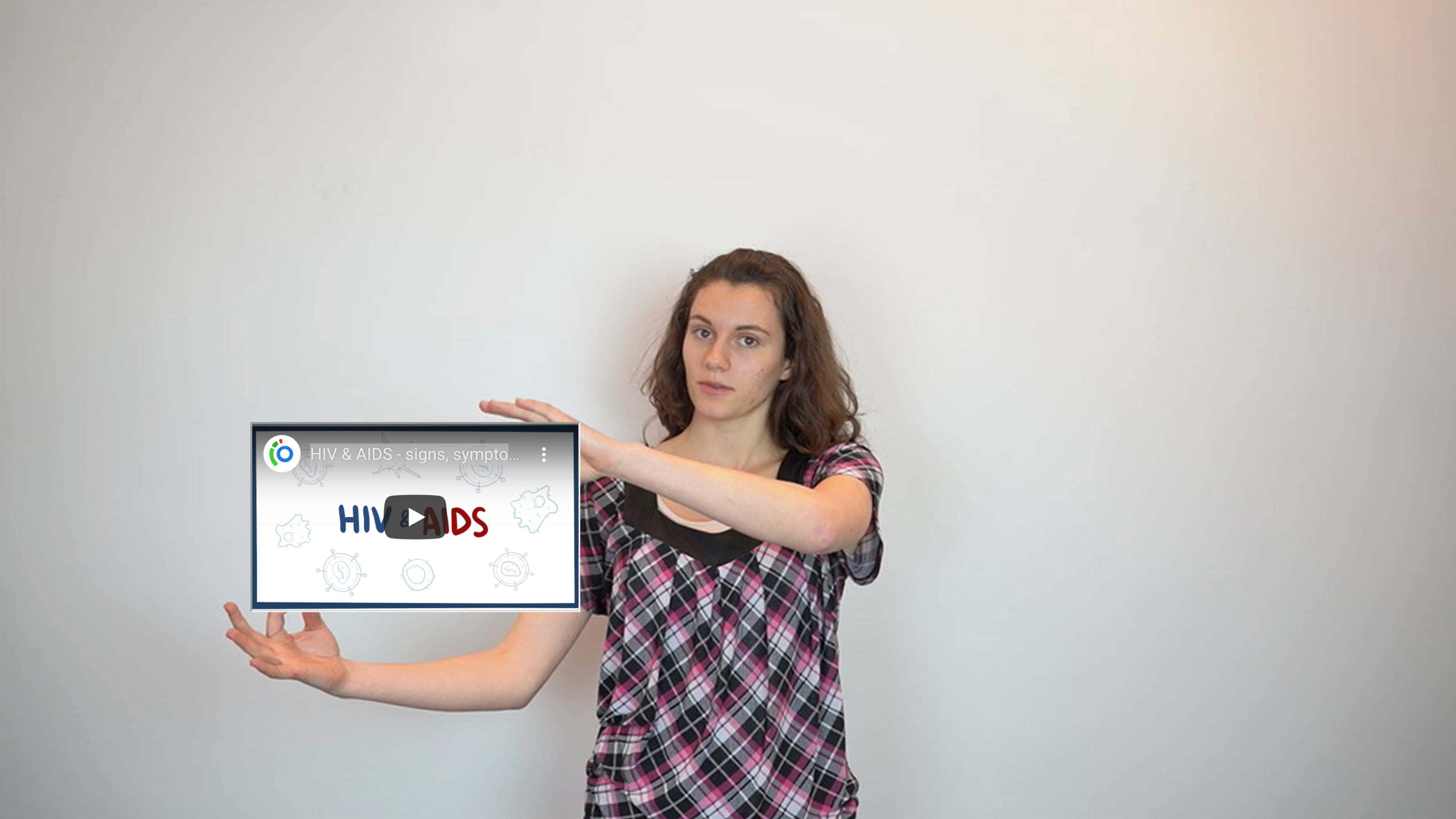}
\includegraphics[width=0.49\linewidth]{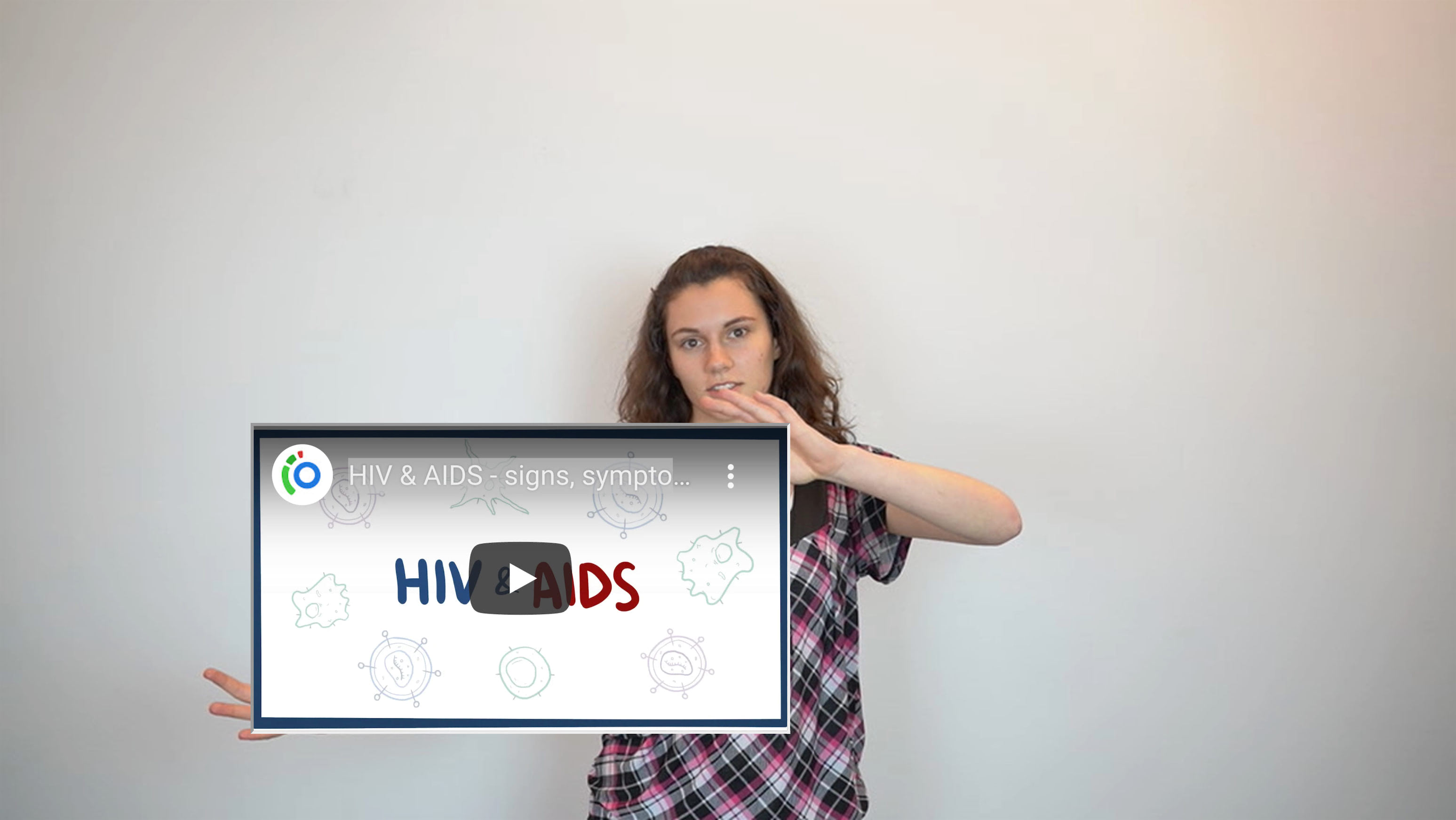}
\caption{Gestural interaction for scaling an object through double finger pinching gesture.}
\label{fig:gesture-scaling}
\end{figure}

\subsubsection{World and Object Tracking}
The system can also understand and recognize the world scene and physical objects. 
By default, the system embeds these textual or visual elements onto the 2D surface in front of the camera, but the system also allows them to embed these elements in the real-world scene. 
For example, the presenter can embed virtual elements in the horizontal or vertical surface. 
In that case, the system can embed the text on the wall, so that even when the camera moves, the text can maintain its position in the 3D scene. 
To specify such a position, the presenter can also use the index finger in the same manner, so that the system detects the position of the vertical surface based on the intersection of the 2D finger position and detected surface, given the raycasting from the 3D camera position. 

\begin{figure}[h!]
\centering
\includegraphics[width=0.49\linewidth]{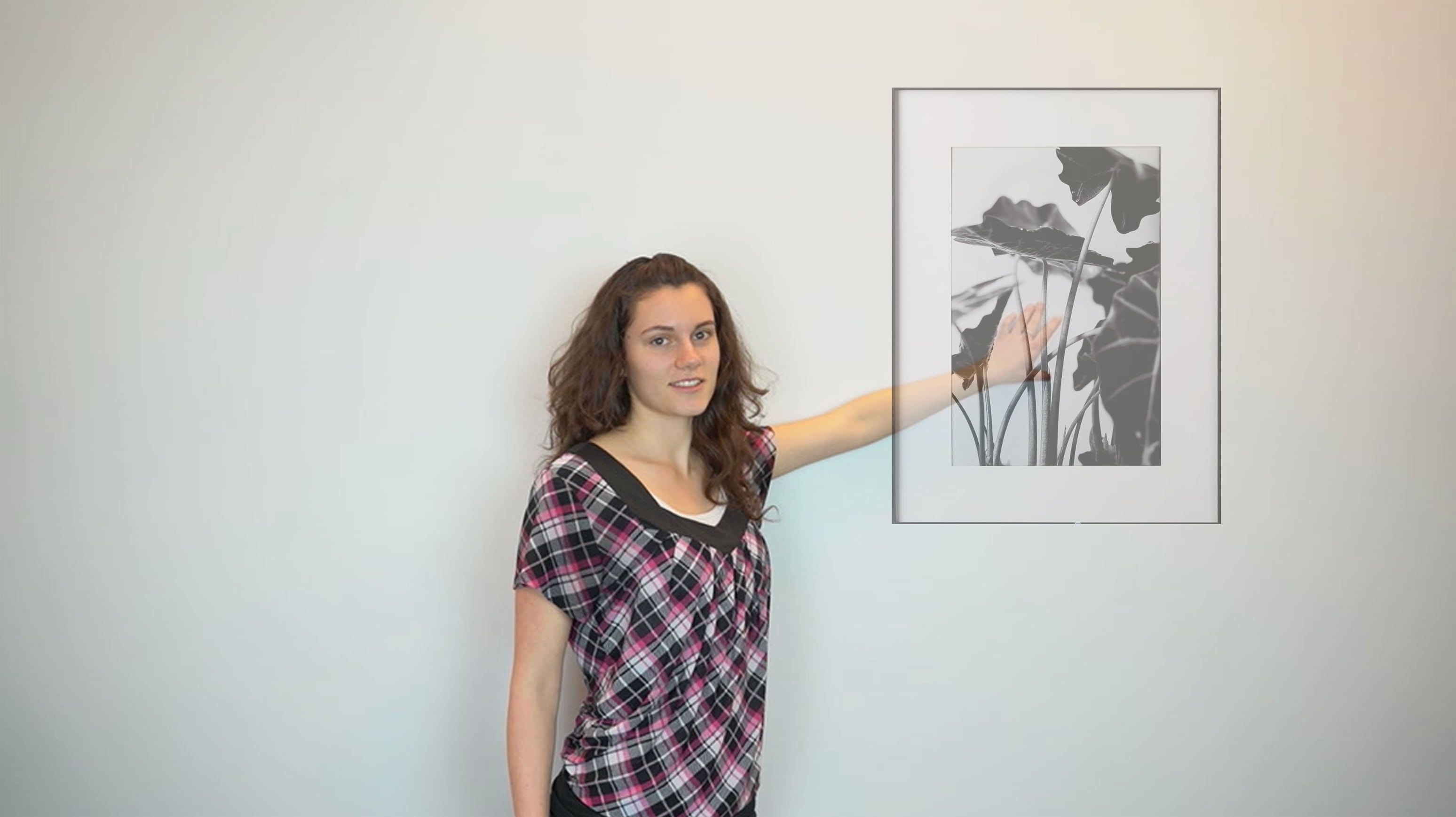}
\includegraphics[width=0.49\linewidth]{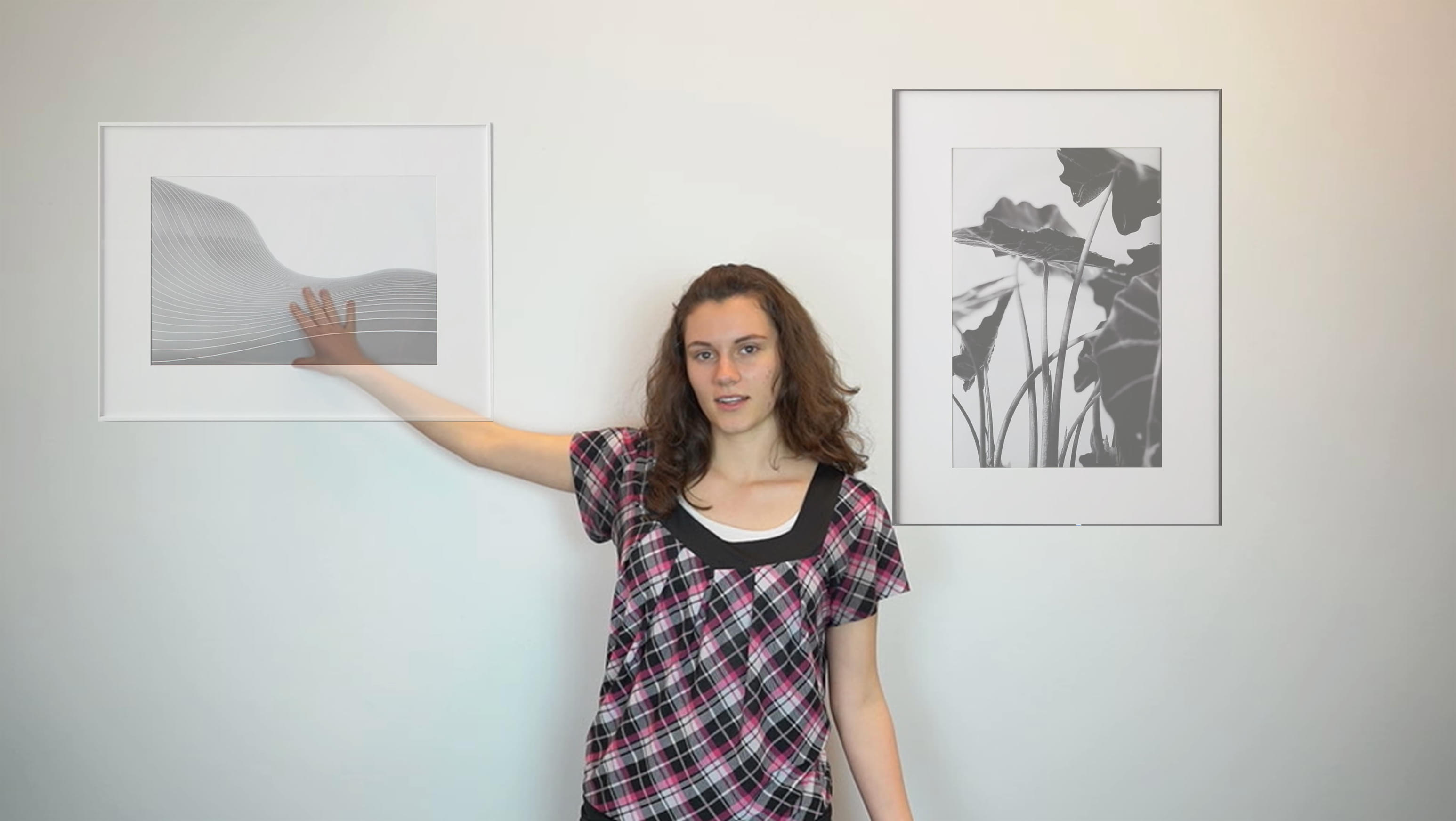}
\caption{Horizontal and vertical surface tracking}
\label{fig:surface}
\end{figure}

The presenter can also embed textual or visual elements associated with a physical object. 
To this end, we employed a simple color tracking technique based on computer vision using OpenCV. 
Therefore, if the user attaches a pre-defined distinct color like a Post-It note to a physical object, then the system recognizes the position of the object with the computer vision color detection.

\begin{figure}[h!]
\centering
\includegraphics[width=0.49\linewidth]{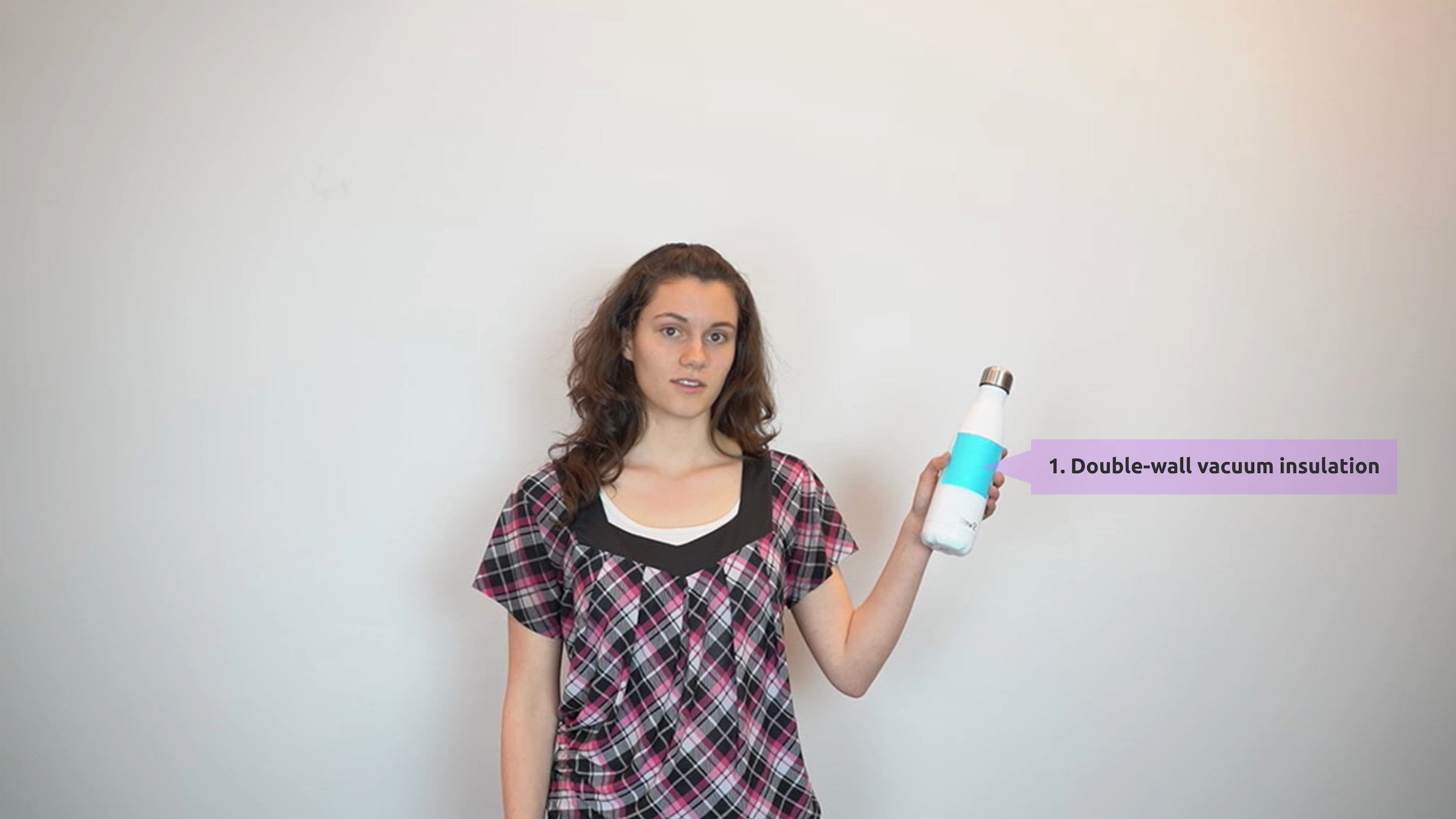}
\includegraphics[width=0.49\linewidth]{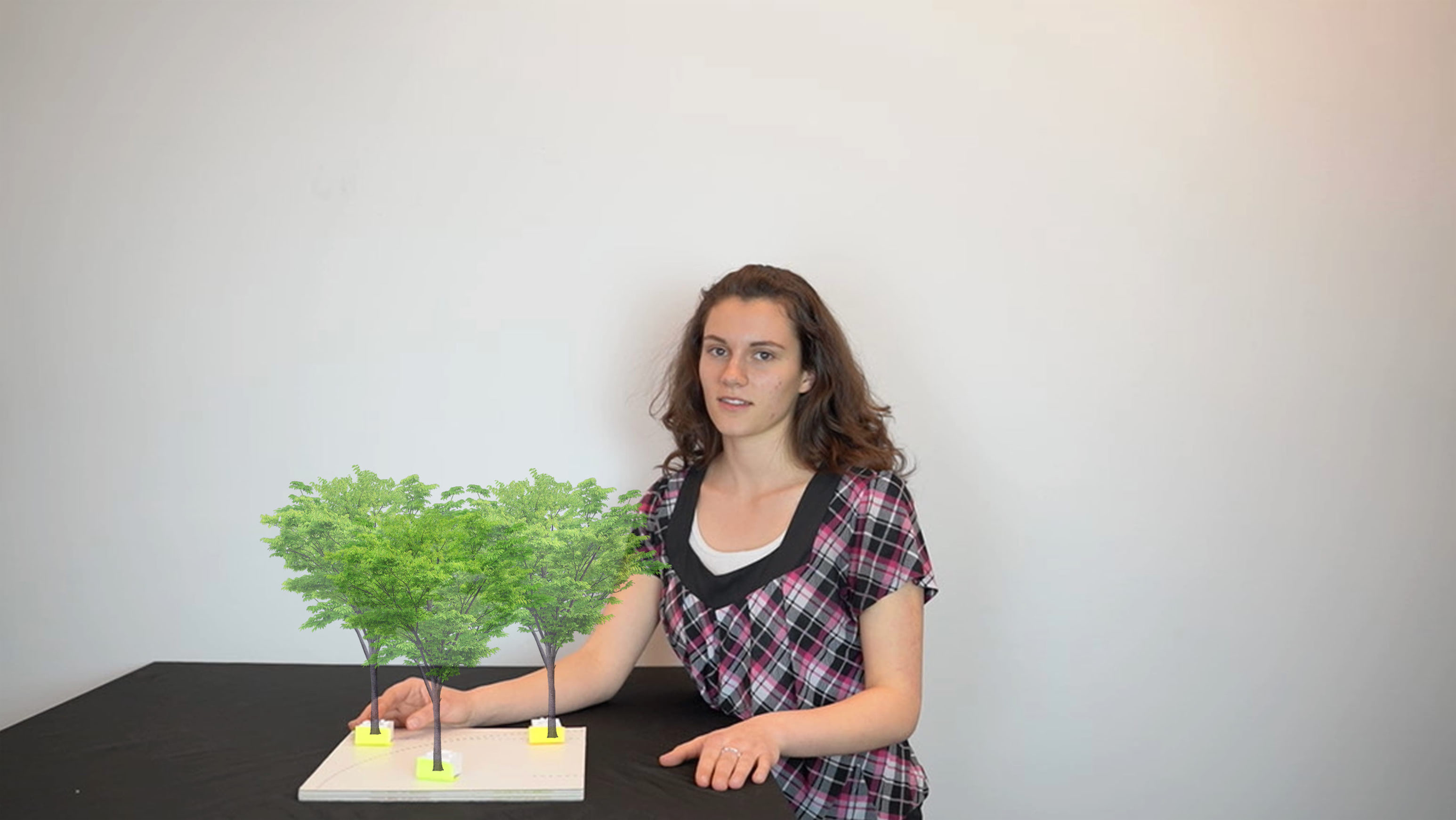}
\caption{Physical object tracking based on colored marker}
\label{fig:color-tracking}
\end{figure}

By default, the system only supports detections of a set of pre-defined colors defined by the authors.
For example, Figure~\ref{fig:color-tracking} illustrates how the system identifies the object based on the pre-defined distinct color markers, such as light blue for left and yellow for right. 
When the system detects the colored marker, then the system automatically embeds the textual or visual elements associated with the object position, so that when the user moves the physical object, the embedded elements also move accordingly, which can enrich the presentation through tangible interaction.

\begin{figure*}[h!]
\centering
\includegraphics[width=0.245\linewidth]{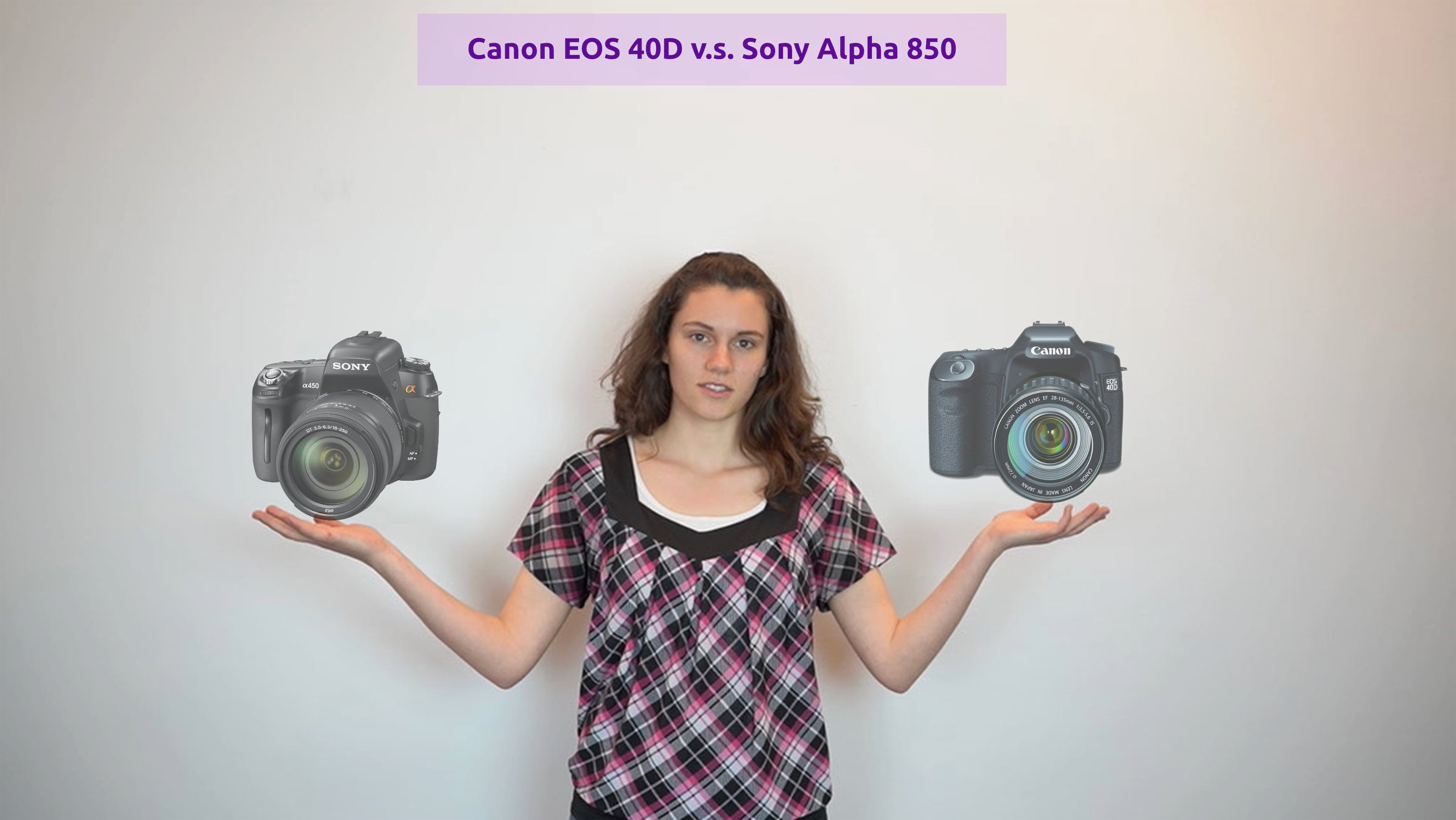}
\includegraphics[width=0.245\linewidth]{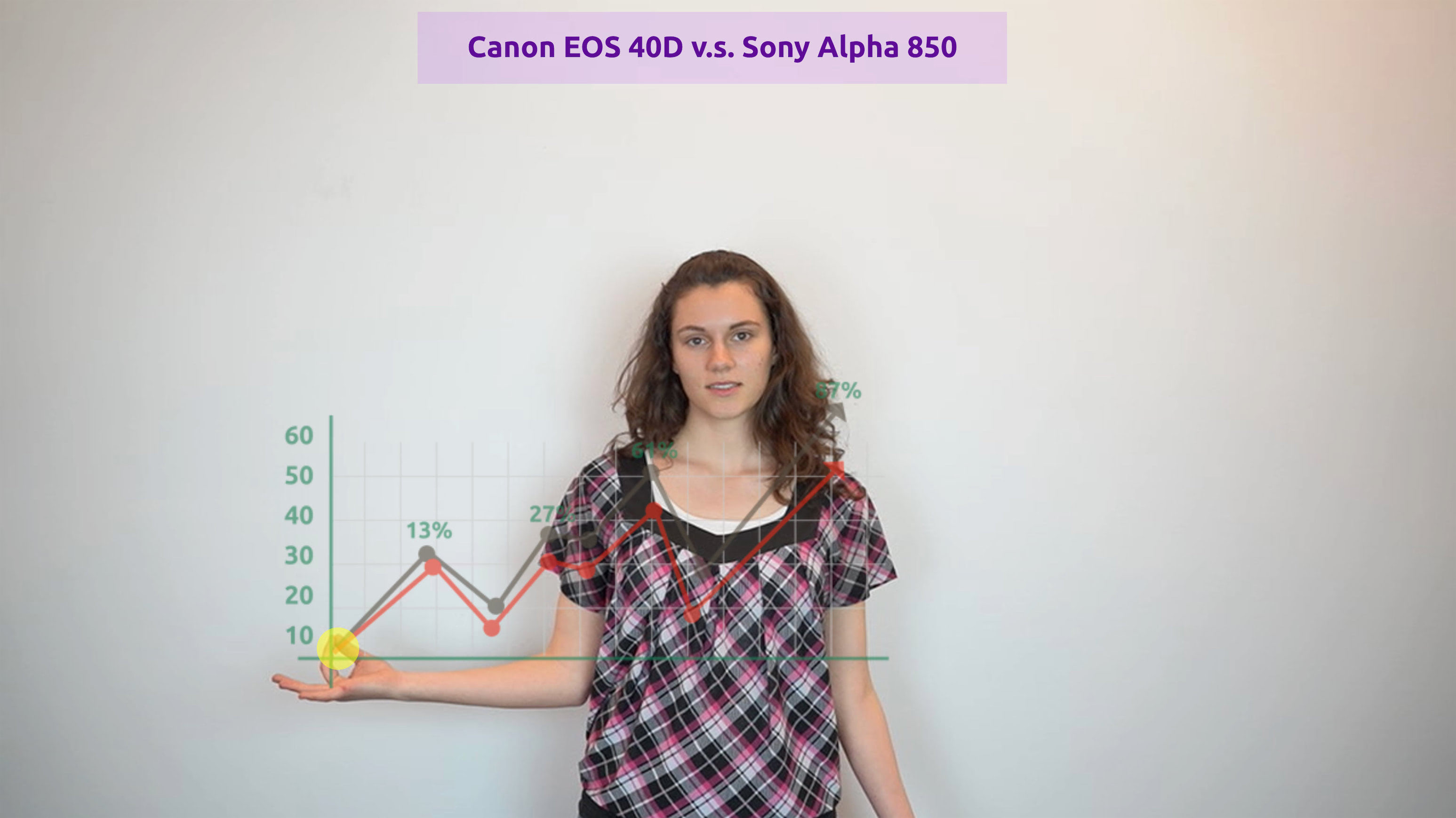}
\includegraphics[width=0.245\linewidth]{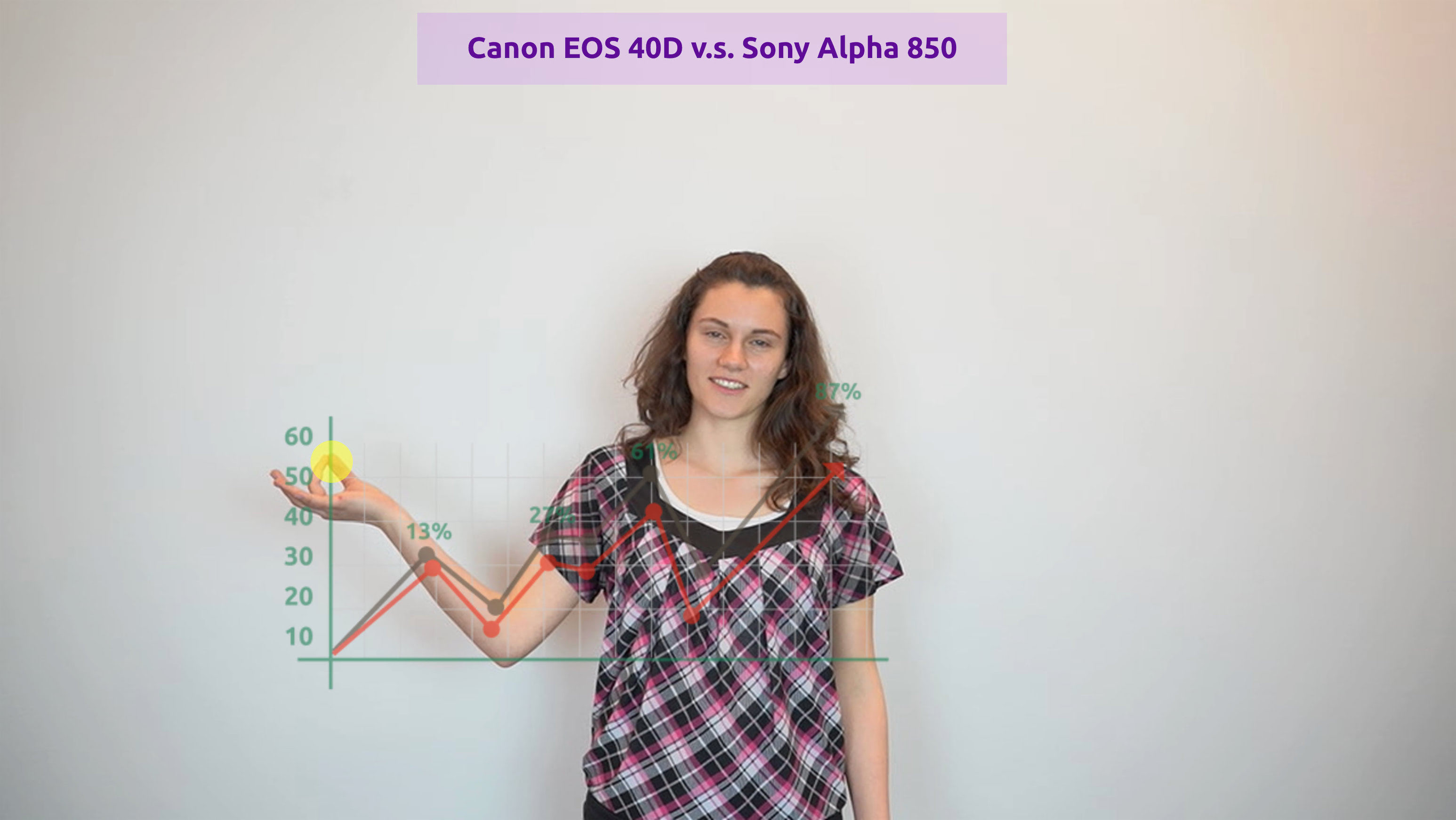}
\includegraphics[width=0.245\linewidth]{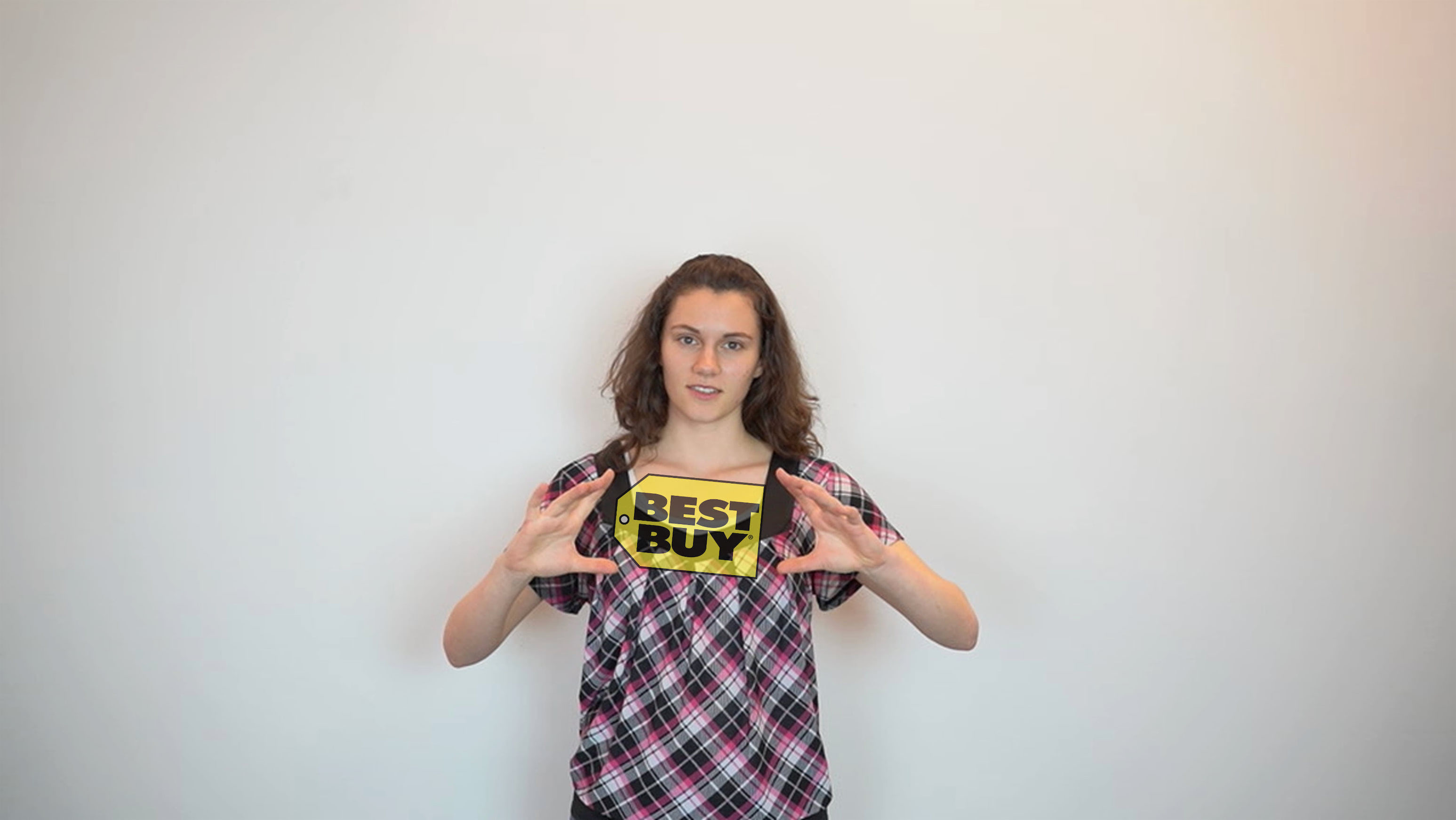}
\caption{Business Meeting Application}
\label{fig:business-meeting}
\end{figure*}

\begin{figure*}[h!]
\centering
\includegraphics[width=0.245\linewidth]{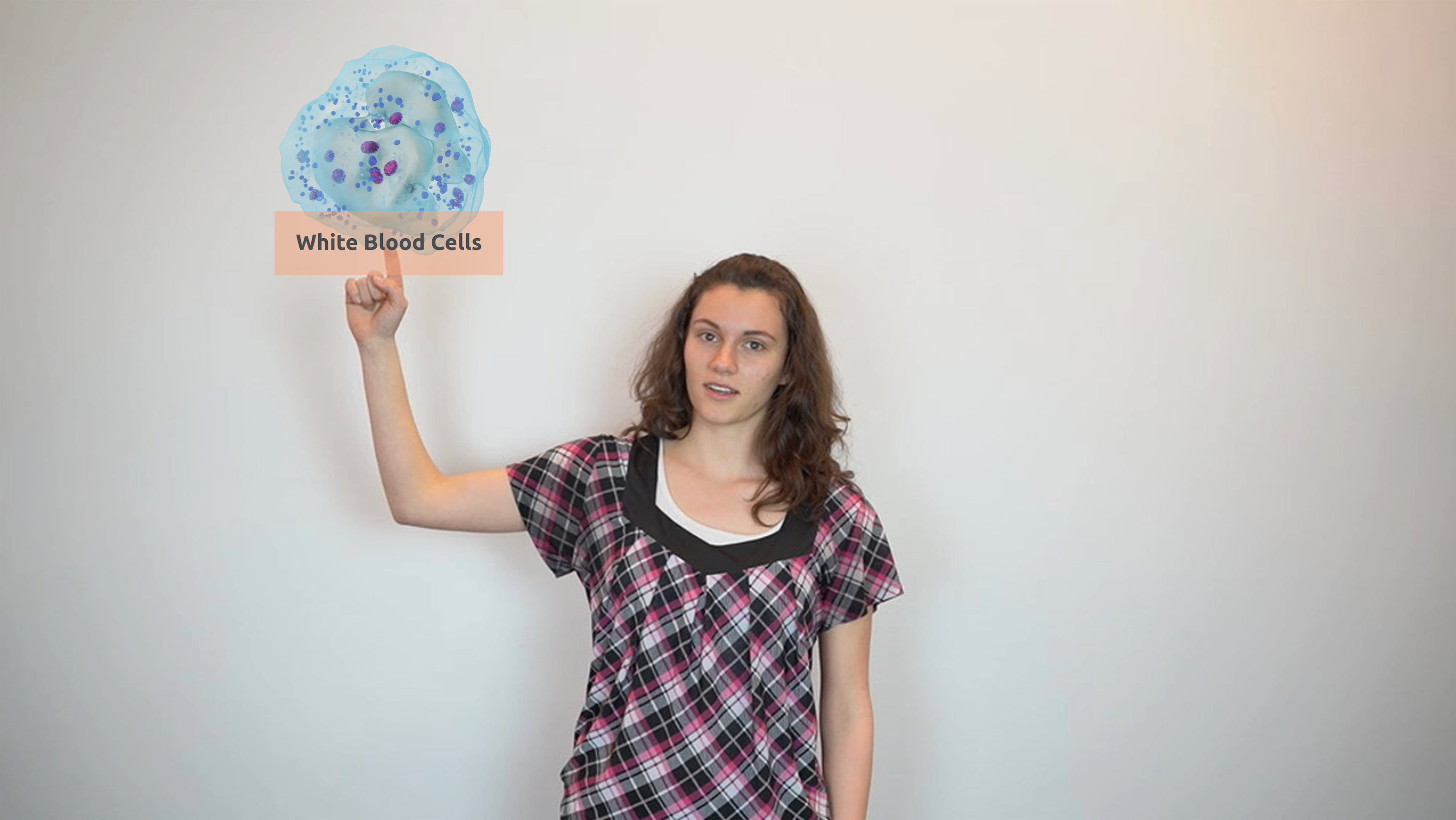}
\includegraphics[width=0.245\linewidth]{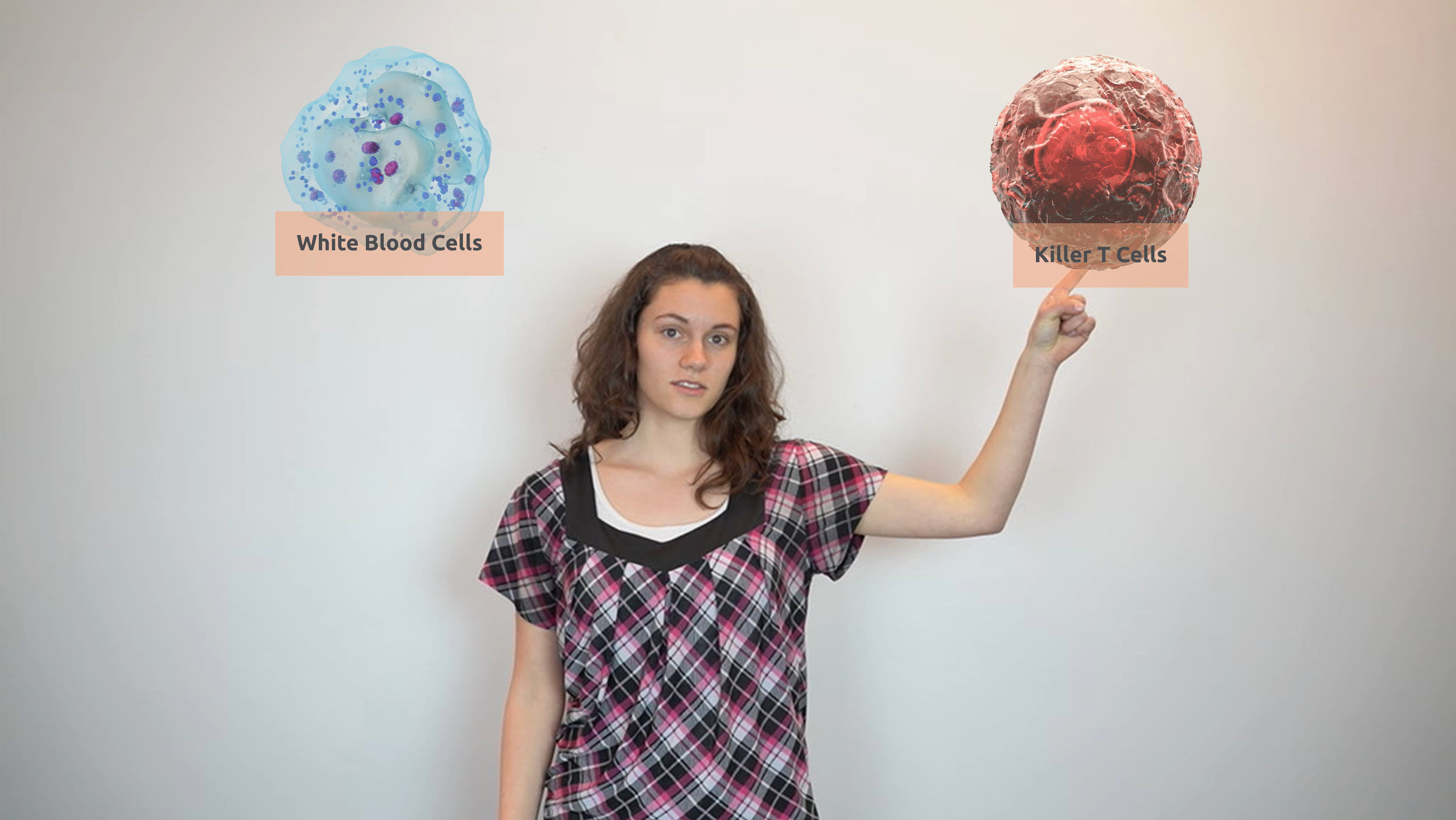}
\includegraphics[width=0.245\linewidth]{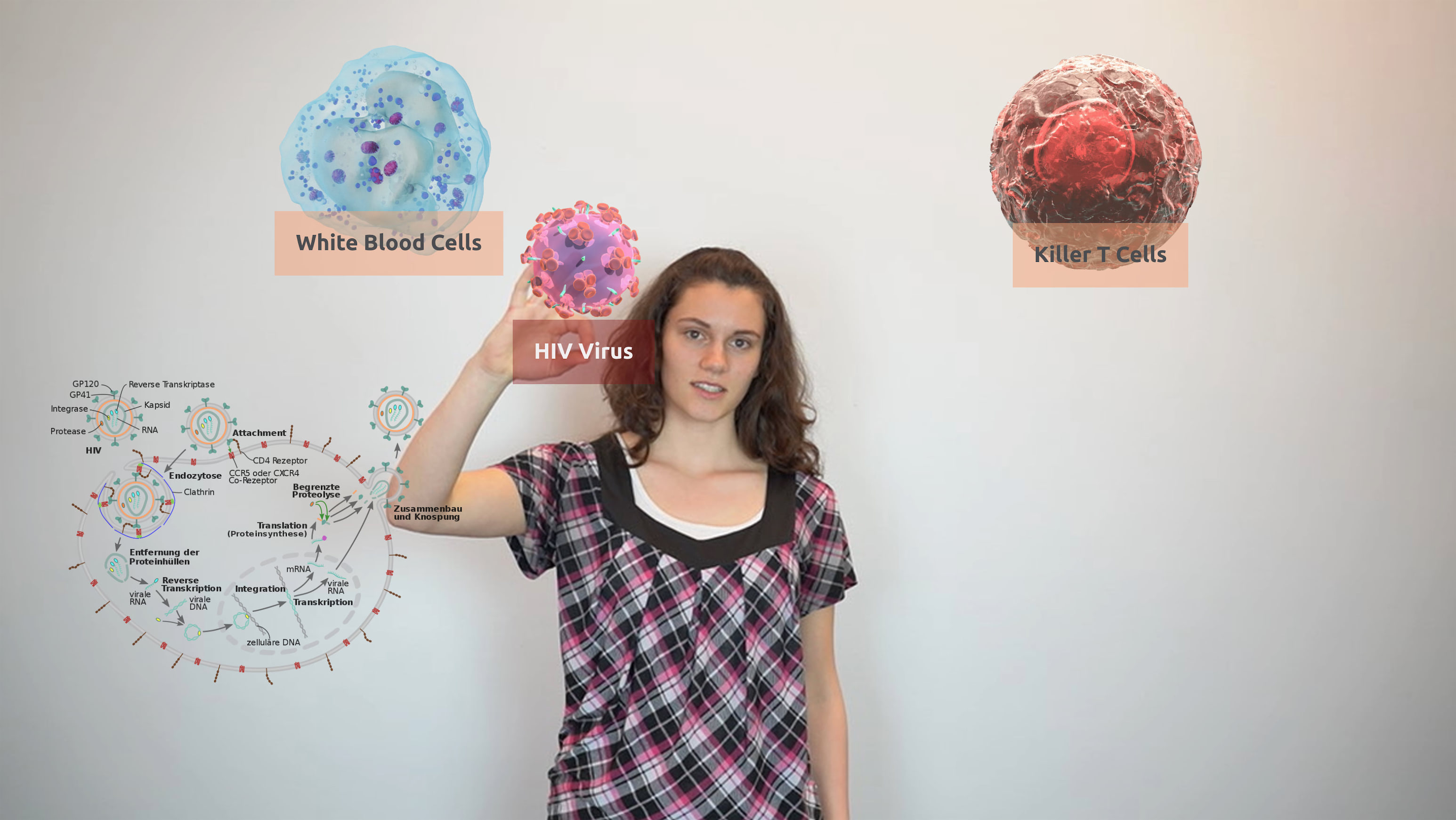}
\includegraphics[width=0.245\linewidth]{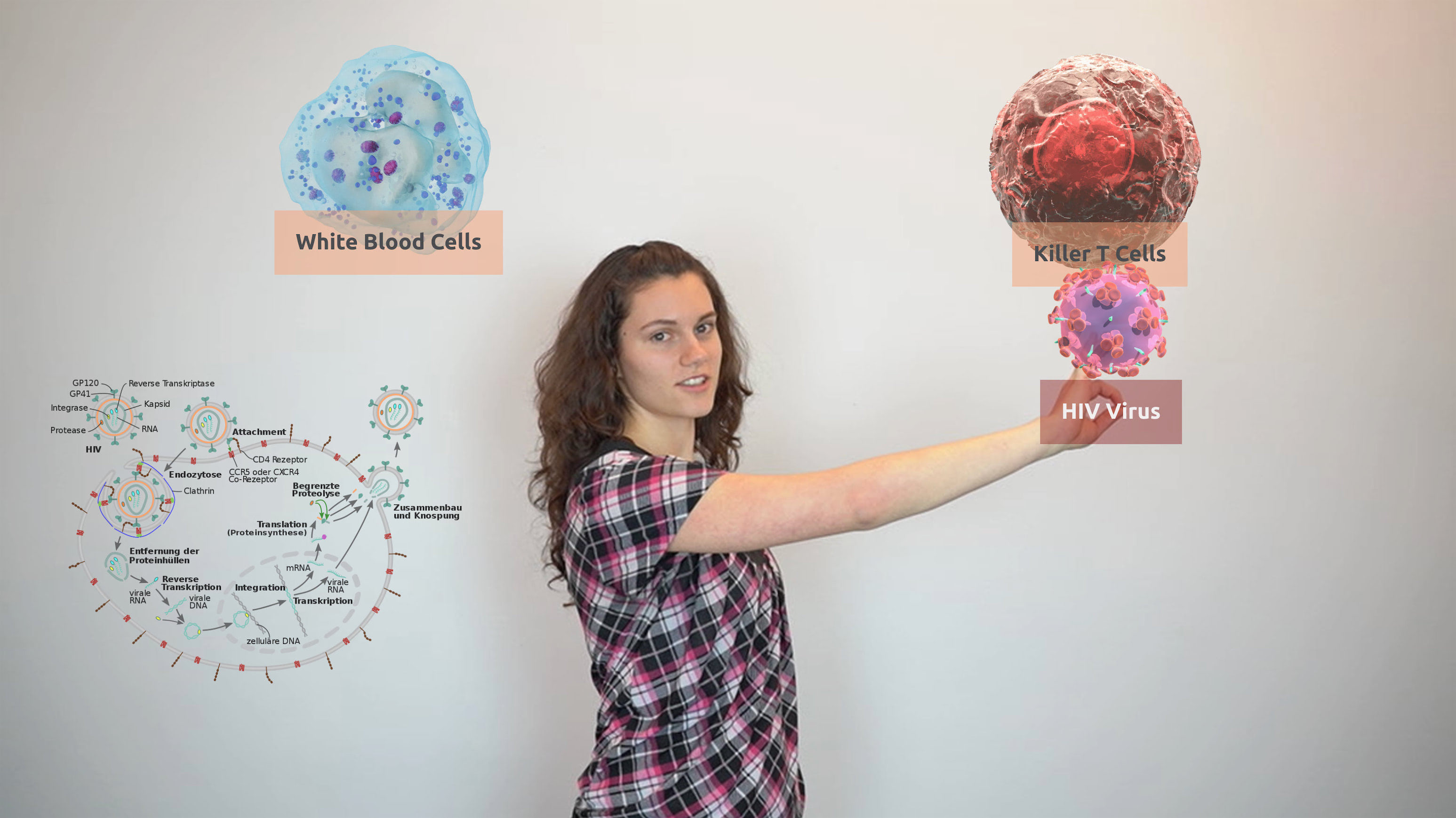}
\caption{Online Lecture Application}
\label{fig:online-lecture}
\end{figure*}

\begin{figure*}[h!]
\centering
\includegraphics[width=0.245\linewidth]{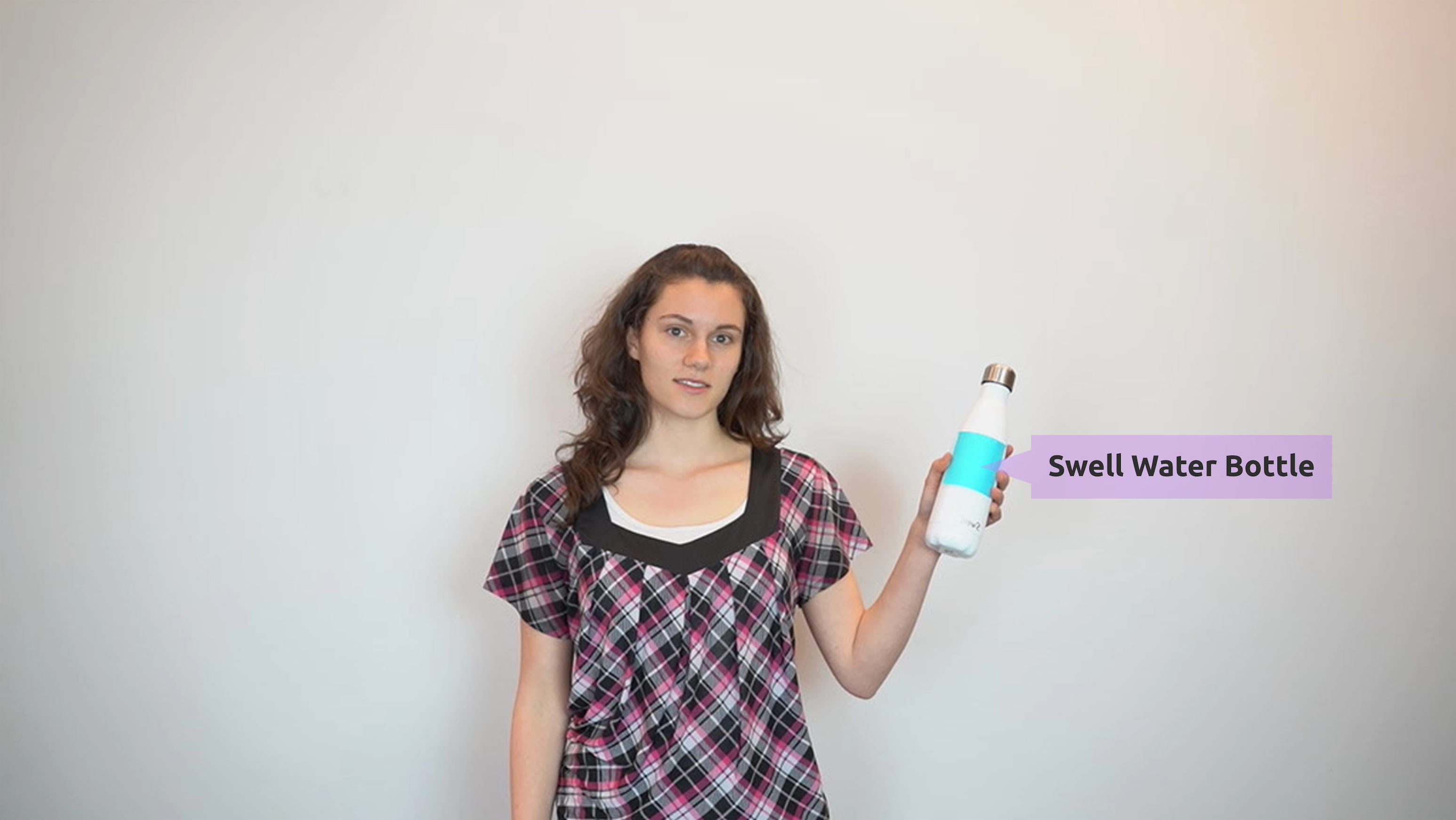}
\includegraphics[width=0.245\linewidth]{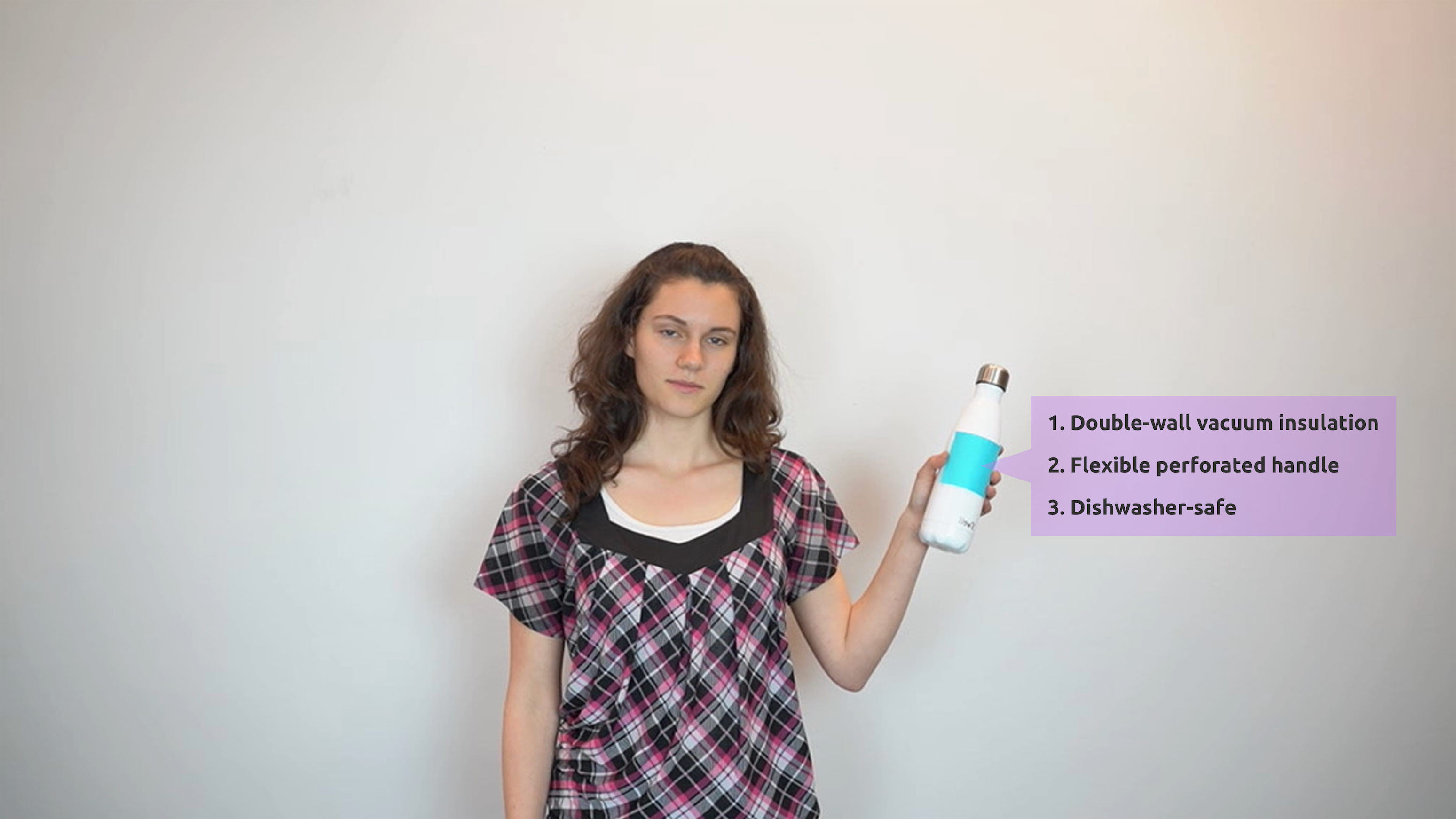}
\includegraphics[width=0.245\linewidth]{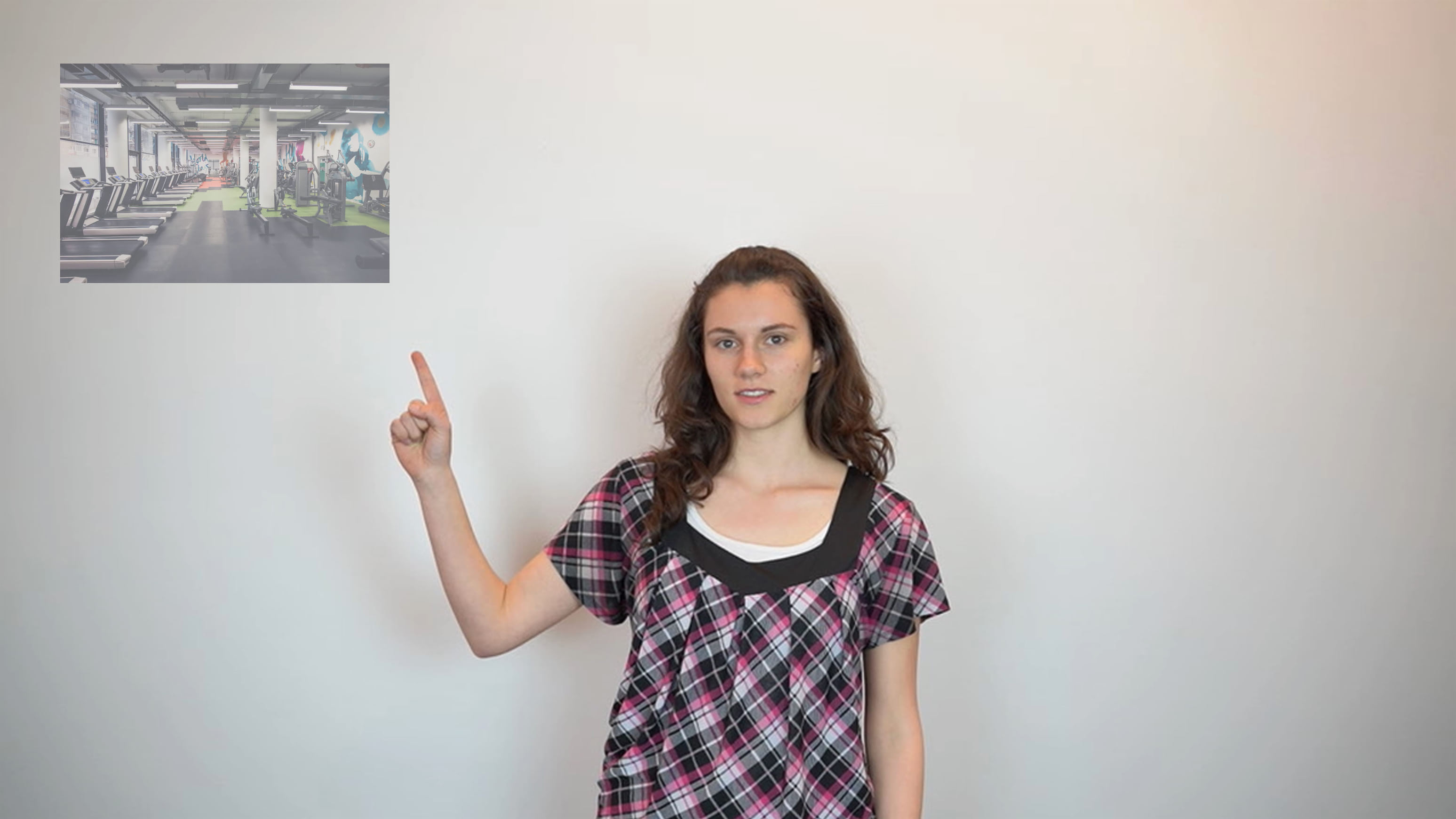}
\includegraphics[width=0.245\linewidth]{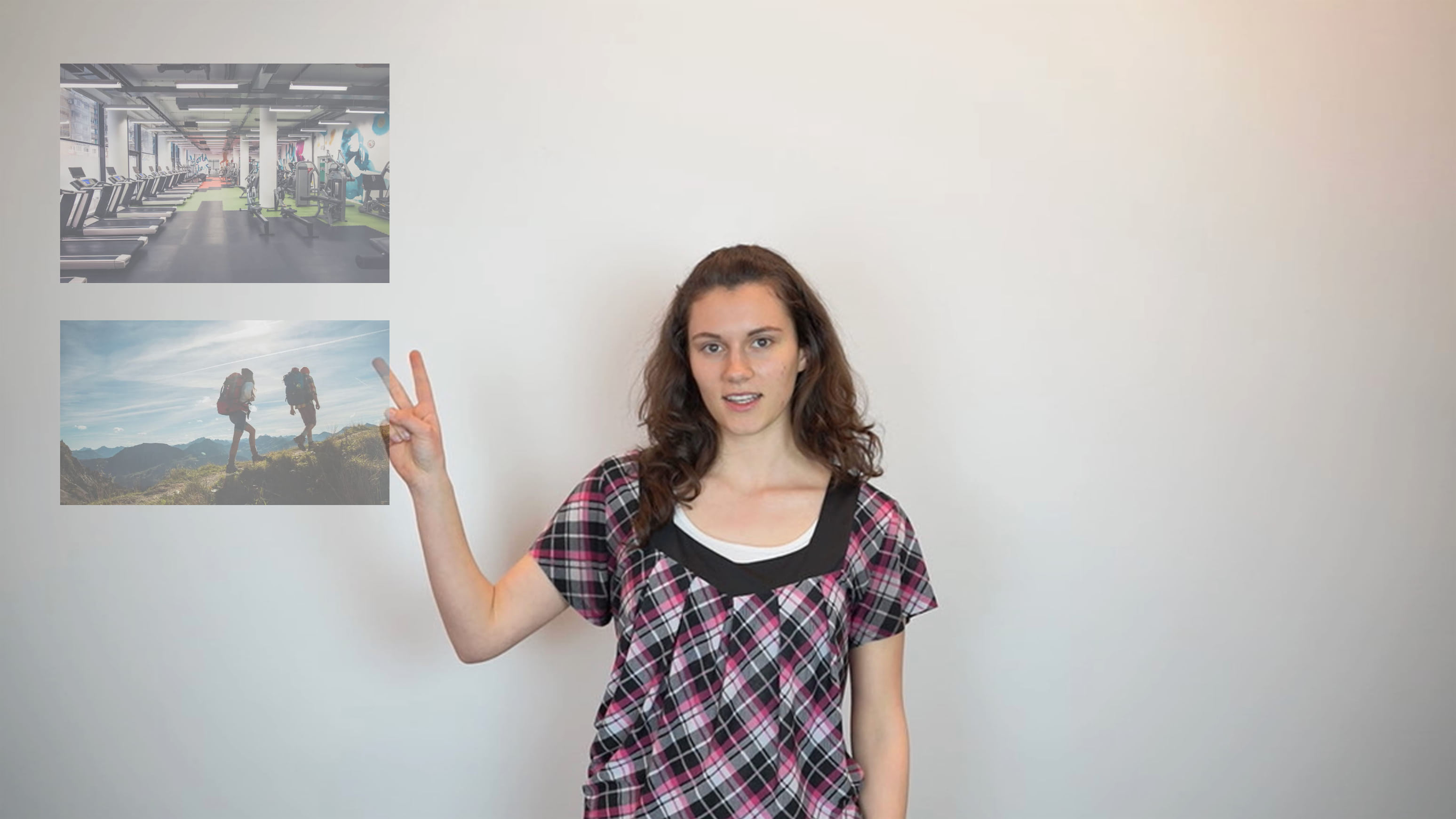}
\caption{E-Commerce Live Stream Application}
\label{fig:e-commerce}
\end{figure*}

\section{Implementation}

\subsection{Speech Recognition}
\system{} uses Google Chrome based Web Speech API to implement asynchronous speech recognition while the presenter is talking, which provides the ability to recognize voice context from an audio input. In our system, we built a speech-to-text pipeline that transcribes the presenter's speech into text transcriptions in real-time, which then is used for \emph{keyword extraction}. 

\subsection{Keyword Extraction}
The transcribed words are sent to Node.js server through WebSocket protocol.
Once the server received the transcribed words, \system{} uses spaCy, a transformer-based NLP engine, to extract the keywords from the speech recognition input. 
In our setting, we use MacBook Pro 2022 14-inch to run the Node.js server and spaCy Python script. 
For the spaCy model, we use the pre-trained model based on the default word embeddings model of en\_core\_web\_sm dataset. 
Through the NLP pipeline, the system detects noun phrases to combine the words and extract keywords. 
Then the system filters the keywords with only nouns, which then are used for \emph{visual embedding and mapping}. 

\subsection{Visual Embedding and Mapping}
\system{} uses Konva.js and A-Frame for the visual embedding of visual elements. 
Since most of the textual and visual elements are 2D, we utilize HTML Canvas as the main rendering method. 
Since A-Frame and Three.js support Canvas texture to show interactive 2D elements as a texture of plane geometry, we leverage this feature for embedded visuals.
Canvas texture also supports HTML embedding such as iframe, so that we can also embed various visual elements such as screens and websites. 
The visual elements are prepared beforehand during preparation using our authoring user interface. From the keyword matching table shown in Figure \ref{fig:system-overview}, the mappings between user-specified keywords and user-defined visuals, such as images, icons, and YouTube videos, allow the presenter to embed the visual elements that match with the extracted keywords. Furthermore, during the keyword matching process shown in Figure \ref{fig:system-overview}, the matched keywords occur with the corresponding visuals whereas the unmatched keywords occur on the screen independently. 
\change{The system also uses simple regular expression and html2canvas to support the template behavior specifically for \textit{list} and \textit{profile} textual elements.} 
Then, the embedded visual elements are supported with a gestural modality for interactions such as specifying the appearance position, dragging and dropping, and scaling. 

\subsection{Gesture Recognition}
Gesture is the supporting modality of our system to create a more engaging augmented presentation. \system{} uses MediaPipe \cite{MediaPipe} for hand landmarks detection.
MediaPipe Hands is a high-fidelity hand and finger tracking solution, which employs machine learning to infer 21 3D landmarks of a hand from just a single frame. With the 21 3D landmarks information, we can further establish a simple gesture recognition system by extracting three basic features: 1) finger position, 2) finger state and, 3) pinching state.

Finger position is the \emph{(x,y,z)} coordinates extracted from the hand landmarks. Finger state indicates the curl state of each finger, namely \emph{Upward} and \emph{Inward}. Upward means that the finger is up, for example, a pointing gesture has the index finger up. Inward indicates that the finger is curled. Pinching state indicates if two fingers are pinching by calculating the \emph{euclidean distance} of two finger positions.


\subsubsection{Pointing Gesture}
Pointing gesture is used for specifying the appearance position of the embedded visual elements. The system detects the pointing gesture by checking the finger state of each finger. When only the index finger is upward and the other fingers are inward, the system will recognize the pointing gesture and retrieve the index fingertip position. Then the index fingertip position is used to specify the appearance position of the visual elements, which produces the \emph{pointing-appearing} visual effects.

\subsubsection{Pinching Gesture}
Pinching gesture is used for interacting with the textual and visual elements. The system checks the pinching state of the index finger and the thumb to detect pinching. Once the pinching gesture is recognized (either left hand or right hand) and the pinching fingers are above a textual or visual element, the presenter would be able to drag and move the virtual object. Furthermore, once the pinching gesture is recognized with both hands, and they are above the textual or visual element, the presenter would be able to scale and rotate the virtual object. The scaling factor is calculated through the \emph{euclidean distance} between two hands, and the rotation degree is calculated through the geometry angle of two hands.

\subsection{World and Object Tracking}
\system{} implements an 8th Wall WebAR pipeline to achieve world tracking, which can recognize 3D surfaces through SLAM and embed textual and visual elements onto the surface in the real world. \system{} also integrates with tracking.js and OpenCV to perform color tracking, which is then used for object tracking by attaching post-it notes or color stickers on the physical objects. 
We detect these colored markers based on a simple RGB value --- once the system detects a certain contour size given a certain range of RGB spectrum, then the system detects and tracks the position of the color.

%% file: 6-application.tex
\section{Applications}

\subsection{Business Presentation}
Our system can be used for presentations such as research pitch or business meeting. 
For example, Figure \ref{fig:business-meeting} shows a \emph{camera product sales presentation} using our system. 
The presenter can start describing the topic based on \emph{Title} text indicating the comparison of the products by saying \textit{``we have two most popular products''}, which is followed by the \emph{associated visuals elements} of two camera pictures that are bound with both hands when saying \textit{``Canon EOS 40D''} and \textit{``Sony Alpha 850''} respectively. Then, the presenter shows the product sales chart over the year.
The presenter shows a \emph{logo visual} between the presenter's both hands when mentioning \textit{``Best Buy''}.

\subsection{Online Lectures}
Our system can also be used for online lectures. For instance, Figure \ref{fig:online-lecture} shows a \emph{biology lecture presentation}. First, the presenter shows the \emph{associated visual} with a \emph{Keyword} text element when mentioning \textit{``white blood cells''} and \textit{``killer T cells''}.
Next, the presenter shows an HIV virus \emph{visual} with \emph{Keyword} while cueing \textit{``HIV virus''}, which is \emph{dragged} and \emph{moved} on the screen approaching the \textit{``white blood cells''} visual through the use of \emph{grasping gesture}. After that, the presenter shows another embedded picture of \textit{``immune system''} while the presenter moves the \emph{HIV virus visual} towards the \emph{killer T cells visual} while demonstrating the HIV invasion process. 

\subsection{E-Commerce Live Streaming}
Virtual selling, remote selling, and social selling are an uprising trend of e-commerce virtual sales that appears on many live-streaming platforms, such as Taobao Live, TikTok Live, and Twitch. Our system can be used for e-commerce live-streaming. To give an example, Figure~\ref{fig:e-commerce} shows a \emph{water bottle virtual sale presentation} using our system. First, the presenter shows the name and brand of the water bottle using \textit{Label and Annotation} and \textit{Object Tracking}. Then, the presenter shows the good features of the water bottle using a template \emph{List} textual element that is associated and bound to the position of the water bottle. 
Next, the presenter shows the \emph{embedded images} while mentioning \textit{``gym and hiking outdoors''}. 
Lastly, the presenter can embed a special discount \emph{icon} and a QR code (Figure~\ref{fig:visual}) for the audience to buy right away during the presentation to deliver a better e-commerce shopping experience.

%% file: 7-user-study.tex
\section{User Evaluation}


\subsection{Method and Study Protocol}
\subsubsection{Participants}
We evaluate our system with 15 participants (9 male, 5 female, 1 non-binary; aged between 18 and 29) from our local community. 
Each participant was compensated with 10 CAD for taking part in the study.
All participants were fluent English speakers (11 native English speakers, and the rest spoke English as their second language) and conducted the study in English.


\subsubsection{Study Protocol} 
Each participant required 30-40 minutes to complete the study as follows:

\subsub{Introduction and Demonstration (10-15 minutes)}
First, the participants were given an overview of the project and then did a practice session to familiarize themselves with the system. 

\subsub{Authoring and Preparation (10-15 minutes)} 
We asked each participant to prepare a 2-3 minutes presentation based on their topic or the topic from the following three: 1) self-introduction, 2) country and family, 3) favorite movie or TV show.
Then, participants were asked to prepare by entering their keywords and associated visuals through the authoring tool. 
We asked the participants to do their best to specifically cover all the system features.

\subsub{Presentation Task (5-10 minutes)} 
Following that, participants went ahead and presented. We did not intervene unless participants asked for aid. No time limit was imposed, but most of the participants performed the presentation for 3-5 minutes. The purpose of this practice was to see if participants could prepare for \change{an engaging augmented presentation with minimal effort using the given system features}. We recorded each participant’s presentation video.

\subsub{Provide Feedback (5-10 minutes)} Finally, participants were asked to fill out a questionnaire via Google Form to provide feedback regarding \system{}'s experience. 
All feedback is anonymized.

\subsection{Results}

\subsubsection{Overall Experiences}
\begin{figure}[h]
\centering
\includegraphics[width=\linewidth]{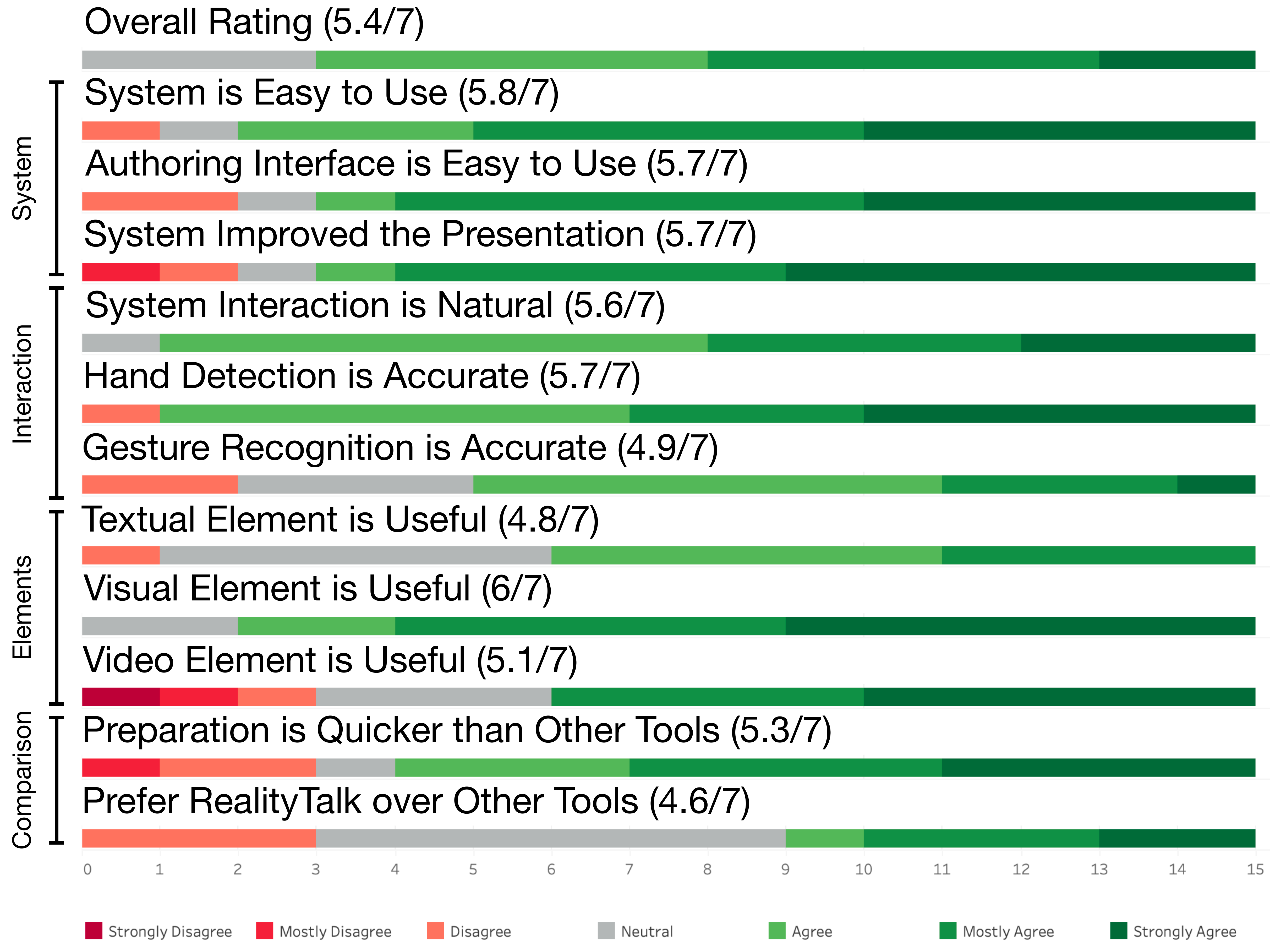}
\caption{User Study Results -- A graph summarizing the 7-point Likert scale responses for 15 participants.}
\label{fig:user-study}
\end{figure}

Figure~\ref{fig:user-study} summarizes the the results.
Overall, the majority of participants responded positively. 
Participants' average rating of \system{}’s overall system was 5.4/7, and they found the system easy to use (5.8/7).
\textit{P15: ``It was very easy to do all the inputting and perform the hand movements myself.''}
Many participants commented that the system is easy (P1, P2, P5, P7, P12) and intuitive (P13, P14, P15).
The participants also found the speech-based augmentation improves their presentations (5.7/7). 
%

\subsubsection{Authoring and Preparation}
The participant also reported positively about the authoring experience (5.7/7). 
\textit{P4: ``It makes preparing for a presentation easier, and you don't need to have a whole set of slides out.''} 
However, one participant found that \system{}’s workflow can add more preparation time.
\textit{P15: ``Having to set up the keywords, objects, and practicing hand gestures would add another step in preparing for a future presentation (aside from putting together the actual presentation).''} 

\subsubsection{Gestural Interaction}
Participants found the system interaction natural (5.6/7) and the hand detection accurate (5.7/7).
\textit{P3: ``the size changing of the pictures worked very nicely. The feature was very helpful, the pictures showed up where I pointed and was smooth.''}
Participants also found the gesture recognition fairly accurate (4.9/7), however, in several instances, the imperfections of gesture recognition hindered the user experience and flow. 
\textit{P6: ``It was a little tricky to grab the text, videos, and images.''}
\textit{P12: ``The words popped up at the tip of my fingers, but were difficult to move around.''}
\textit{P2: ``The scaling feature is a little too sensitive, but is responsive when it is supposed to be.''}
\change{From the recorded user study videos, we also noticed that while the pointing gesture is the most accurate, pinching and scaling gestures sometimes are too sensitive.}


\subsubsection{Usefulness of Each Element}
\change{For the comparison of each element, participants gave the ratings of the usefulness as follows: textual elements (4.8/7), video (5.1/7), images and icons (6.0/7).
The majority of the participants found the speech-driven visual elements the most helpful.} 
\textit{P8: ``Showing a video by saying the keywords was really engaging the presentation.''} 
\change{On the other hand, due to the recognition error of the ML pipelines, especially the speech recognition with the WebSpeech API, participants often struggled to trigger the corresponding textual or visual elements as expected.
\textit{P15: ``The system was not understanding my words easily, but that could be due to my accent.''} 
\textit{P1: ``Speech was clear, but some pictures refused to pop up, making the overall experience somewhat stressful.''}} 
In addition, the duplicated textual and visual elements could lead to a convoluted screen.
\textit{P10: ``If the keyword is supposed to be something general or mentioned several times in a few seconds, it shows up too many images and distracts me. I think it is better to make it not pop up the same keyword or images if it is still on the screen.''} 
\change{These recognition errors introduce a great amount of friction and distraction, which requires future work.
}


\subsubsection{Comparison with Other Practices and Tools}
\change{While some participants liked our system over other tools (\textit{P8: ``It will be better than PowerPoint.''}), the participants provided the lowest rating (4.6/7), when asked if they would prefer \system{} over traditional slide-based tools.
We also acknowledge that \system{} is not intended to replace or outperform traditional tools (e.g., PowerPoint, Keynote, OBS), but rather to propose an alternative approach.}
In fact, participants think the preparation is quicker than the other tools (5.3/7) and they appreciate the improvised, non-linear aspect of \system{} (P7, P14, P15), which provides a unique benefit compared to slide-based practices.
For example, P7 described that \system{} is preferred in a more dynamic and less structured setting for improvisational talk, as opposed to lecture-style presentations.
\change{Therefore, we think while traditional tools like Keynote or OBS may produce higher-quality outcomes when given abundant preparation time, \system{} enables quick and improvised presentations with minimal or even no preparation time.}

\subsubsection{Potential Use Scenarios and Applications}
In terms of the potential scenarios, participants mention that live-streaming would be one of the most promising ones.
\textit{P4: ``Streaming on platforms like Twitch could really use this. It could join Steamdecks in making the streaming experience easier.''} 
\change{Based on the response, the most relevant use scenarios could be the one that requires improvised and non-linear practice, such as semi-structured interviews and webinars.}
P1, P3, and P10 also pointed out how visual dominant presentations such as TED Talks could be enhanced using \system{}.
They noted, however, the system needs to be more reliable for formal presentations such as lectures and business presentations.


\subsubsection{\change{Objective Measurements}}
\change{Finally, we also report the objective measurements based on the analysis of video recording (the average recorded video length is 123.57 seconds, SD=36.55).}

\subsub{Hand Interaction Error}
\change{lost hand tracking (M=2.36 times, SD=2.02), pointing (M=0.88 times, SD=0.99), scaling (M=1.77 times, SD=1.30), and pinching (M=3.64 times, SD=2.31).} 

\subsub{Speech Recognition Error} 
\change{unrecognizable keywords (M=0.5 times, SD=0.65), incorrect keywords (M=2.79 times, SD=2.46).}

\subsub{Textual and Visual Elements Error} 
\change{keyword recognized but failed to show visuals (M=0.36 times, SD=0.63), incorrect position of the widgets (M=1.07 times, SD=0.83), widgets cover user (M=1.14 times, SD=0.86), duplicated visuals (M=1.29 times, SD=1.38).}

\subsub{Latency of Appearance} 
\change{The latency between the user's utterance and the visual appearance is 1.47 seconds (SD=0.23), and the average keyword utterance is 2.07 times (SD=1.98).}

\change{Since our system is only a proof-of-concept prototype, we notice there were quite a few errors in each participant's recorded video. These errors, which mostly come from the error of the ML pipeline, indicated that the system requires more work for real-world usage. But we also noticed that some errors could be avoided if the participant were more familiar with the system features. For example, the hand interaction error rate varies from 0 to 9 times and some participants' performances are smoother than the others. That is because some participants did not know how to use the gesture appropriately after they tried and failed multiple times. (e.g., pinching gesture to move the virtual elements). We believe if they were given a few more practices this error could be avoided.}

%% file: 8-future-work.tex
\section{Limitations and Future Work}

For future work, we plan to integrate the suggestions from the participants for more reliable systems. On top of that, we discuss future research opportunities for further exploration. 

\subsection{Semi-Automated Suggestions}
\change{Although it is minimal, the current system still requires manual preparation, which makes some participants feel the process is tedious (P9).
More importantly, the user needs to think in advance, as the system cannot show any visuals if keywords are not provided beforehand.
In the future, we could tackle this problem by leveraging semi-automated suggestions.
As mentioned, the naive automated approach (e.g., automatically show images extracted from Google Images) does not work well, but the user could scope down the keywords or images with semi-automated approach (e.g., only using keywords listed in an index of a textbook for a classroom lecture; only using images in specific papers for a panel discussion at a research conference).
In this way, the future system could minimize the preparation time and make it more improvisational.}

\subsection{Customizability of the Presentation}
The participants stated that current video editing tools can achieve a higher degree of flexibility and customizability than \system{} (P2, P4, P12). 
\change{We aim to further improve our system by achieving higher accuracy and flexibility. For example, a possible extension to the authoring interface could be a custom template for various situations like Powerpoint or Keynote.
In addition, the flexibility of keyword mapping should be improved as the current system only supports one-to-one keyword mapping, which might limit users' creative freedom. 
These flexible customization for better quality presentations is important for a future improvement.}

\subsection{Deployment in the Wild}
On top of these improvements, we are interested in deploying our system in the-wild. 
For example, it is interesting to investigate whether our tool is good enough for video production for YouTuber or not.
Alternatively, we are also interested in deploying our system for specific use cases like classroom lectures.
\change{In such cases, the audience aspect should be also considered, as our evaluation did not consider a comprehensive audience's experience.
Future work should also investigate how presenters and audiences interact with each other, such as how they interrupt the presenter, ask questions, and discuss things during the presentation.}

\subsection{Presentation with HMDs or Spatial AR}
For the current prototype, our system focuses on the webcam on a desktop PC or smartphone for the availability of the hardware.
But, it is also interesting to see how our system works with mixed reality headsets. 
We expect this scenario will improve some gestural interaction as these headsets have more accurate hand and object tracking.
In addition, this mixed reality environment allows us to embed these textual and visual elements in 3D space, which provides a more immersive experience for both presenter and audience. 
Alternatively, our system can also be integrated for spatial AR.
For example, our system could use projection mapping like HoloBoard~\cite{holoboard} for immersive classroom presentations and lectures.

\subsection{Augmented Conversation and Discussion}
Finally, we also see the great potential of this approach (i.e., the combination of NLP and AR) beyond presentation.
For example, we could enhance various modes of communication such as everyday conversation, brainstorming discussion, and interaction with smart speakers.
In these situations, it requires the different requirements (e.g., cannot prepare beforehand), design and implementation constraints (e.g., using HMDs instead of webcam and screens), and evaluation criteria (e.g., the interface needs to be less distracting).
For future work, we will investigate how the speech-driven augmentation can be used in these other domains. 

%% file: 9-conclusion.tex
\section{Conclusion}
We present \system{}, a system that augments live video presentation with speech-driven interactive virtual elements. 
\system{} enables the users to create these augmented presentations \textit{in real-time} by embedding interactive textual and visual effects in AR.
The presenter can show and manipulate these effects interactively based on real-time speech and gesture interactions.
Based on our analysis of existing videos, we propose a novel set of interaction techniques for \system{}. 
Our study confirms that our approach delivers a unique improvisational experience for presentations.
The evaluation of our system points to the future promise and potential of this system.